\documentclass[12pt,fleqn]{article}

\usepackage{graphicx}
\usepackage{comment}
\usepackage{amsmath}
\usepackage{amssymb}
\usepackage{enumitem}
\usepackage[tight]{subfigure}\subfigtopskip=0pt

\numberwithin{equation}{section}

\usepackage{pslatex}

\makeatletter
\def\section{\@startsection
 {section}{1}{0mm}
 {-\baselineskip}
 {0.5\baselineskip}
 {\bf}}
\def\subsection{\@startsection
 {subsection}{2}{32pt}
 {-0pt}
 {-\baselineskip}
 {\bf}}
\def\subsubsection{\@startsection
 {subsubsection}{3}{32pt}
 {-0pt}
 {-\baselineskip}
 {\itshape}}
\makeatother

\begin{document}

\noindent
\begin{center}
{\bf Simulation of Fast Magnetic Reconnection using
  a Two-Fluid Model of Collisionless Pair Plasma without
  Anomalous Resistivity}\footnote{Appeared in {\em Proceedings of
  the 19th Annual Wisconsin Space Conference (2009).}}

E. Alec Johnson
(ejohnson@math.wisc.edu)\footnote{This research was supported by
a Wisconsin Space Grant Consortium Graduate Fellowship
for 2008-2009.}
\\
James A. Rossmanith
(rossmani@math.wisc.edu)


Department of Mathematics\\
UW-Madison
\end{center}

\begin{abstract}
For the first time to our knowledge,
we demonstrate fast magnetic reconnection near a magnetic
null point in a fluid model of collisionless
pair plasma without resorting
to the contrivance of anomalous resistivity.
In particular, we demonstrate that fast reconnection
occurs in an anisotropic adiabatic
two-fluid model of collisionless pair plasma with relaxation toward
isotropy for a broad
range of isotropization rates. For very rapid isotropization
we see fast reconnection,
but instabilities eventually arise that cause numerical
error and cast doubt on the simulated behavior.
\end{abstract}

\section{Overview}

\subsection{Motivating Problem.}

We have been working to develop algorithms that efficiently
model fast magnetic reconnection in collisionless space plasmas.
A \emph{plasma} is a gas of charged particles and the accompanying
magnetic field that it carries. The ability to simulate plasmas
efficiently over a wide range of phenomena and scales is
essential to understanding and predicting the behavior both of space
plasmas and of industrial fusion plasmas.

The two basic types of plasma models are kinetic models and
fluid models.
Fluid models are computationally less expensive attempts to approximate
kinetic models.
Kinetic models represent particles or evolve the
space-velocity distribution of particles. Fluid models evolve
moment averages (e.g.\ density, momentum, or energy) of assumed
(e.g.\ Maxwellian) velocity distributions.  
Fluid models give good accuracy for highly collisional plasmas.
For rarefied space plasmas, however, particularly near magnetic
null points, the regular velocity distributions assumed by fluid
models often fail to hold, since the particles move essentially
unconstrained.
%

The phenomenon for which fluid models of plasma have
been most apt to fail is collisionless fast magnetic reconnection.
It is precisely this phenomenon 
which has proved most critical
to understanding and predicting the volatile dynamics
of astrophysical plasmas, including solar storms and 
geomagnetic substorms in Earth's magnetosphere.
The critical physical role of fast reconnection and the failure
of fluid models to capture it has prompted extensive studies
using particle-based simulations of collisionless magnetic
reconnection. \emph{Our objective is to study the ability of
fluid models to match particle-based simulations of collisionless
fast magnetic reconnection.} We have concentrated our effort
specifically on fast magnetic reconnection in collisionless
\emph{pair} plasmas.
A \emph{pair plasma} is a plasma whose positively and negatively
charged particles have the same charge-to-mass ratio.
The physical example of a pair plasma is an electron-positron plasma,
of interest to astrophysicists.
%

\subsection{Historical development.}

The GEM magnetic reconnection challenge problem
\cite{article:GEM} identified Hall effects as critical to fast
magnetic reconnection in electron-ion plasmas. Since Hall
effects are absent for electron-positron (pair) plasmas, this
prompted Bessho and Bhattacharjee \cite{article:BeBh05, article:BeBh07}
to demonstrate
via particle simulations that fast magnetic reconnection occurs
even in collisionless pair plasma, which they attributed to pressure
anisotropy.

The next challenge was to demonstrate fast reconnection in a
fluid model of pair plasma.
Assuming the ubiquitous presence of a strong
background magnetic guide field (which constrains charged
particles to move in tight spirals)
allowed Chac\'on et al. \cite{article:ChSiLuZo08} to
develop an analytical fluid theory of fast reconnection in
magnetized pair plasma.

For the case where there is a magnetic null point,
Zenitani et al. 
\cite{article:ZeHeKl09} demonstrated fast reconnection in a
two-fluid model of relativistic isotropic pair plasma.
Their model assumes a spatially dependent anomalous resistivity,
as has been used with resistive single-fluid MHD to simulate fast
reconnection.
By selecting a resistivity
with an anomalously high value near the X-point,
one can essentially prescribe the desired rate of reconnection
(as determined from PIC simulations);
the elusive goal is to find a simple and generic expression 
for anomalous resistivity that works for a broad range of
problem conditions.
We remark that resistive single-fluid MHD does not
assume any particular ratio of mass-to-charge ratios between the two
species and thus (with an appropriate choice of anomalous
resistivity) could be used as a fluid model of pair plasma.

\subsection{Our work.}

Rather than resort to an anomalous resistivity, we
seek generic two-fluid moment closures that
give reconnection behavior in agreement with particle-based
simulations.
We have studied reconnection in
five-moment (isotropic)
and ten-moment (anisotropic) adiabatic fluid
models of collisionless pair plasma with varying
rates of relaxation toward isotropy.
We implemented conservative
shock-capturing
Discontinuous Galerkin
two-fluid five-moment and ten-moment plasma
models, following Hakim et al. \cite{article:HaLoSh06,article:Ha08}.

We initially adopted the modified GEM settings of
\cite{article:BeBh05, article:BeBh07}, but 
found that for pair plasmas the large aspect ratio
of the reconnection region gives rise to
a secondary instability, namely, the unpredictable formation of
magnetic islands, making it difficult to obtain
demonstrably converged results, as seen in our paper,
\cite{article:JoRo08a}.
Essentially, for the pair plasma case of the GEM problem,
in contrast to the electron-proton case,
the tearing instability wants to produce smaller magnetic
islands; since the GEM problem \emph{is} the formation of
one big magnetic island, we can avoid the instability by
reducing the size of the domain.
Pair plasma involves no need to resolve scale separation
between species, so arguably a smaller domain is acceptable.
Therefore, to eliminate the secondary instability and
to reduce computational expense
we multiplied the dimensions of the domain by one half.

We also chose to focus on the case where both species have the
same temperature. In this case there is complete symmetry between
the two species, and the number of equations needed is halved. In
the case of zero guide field the GEM problem is symmetric about
both the horizontal and vertical axes. We enforced all these
symmetries, reducing computational expense by a factor of eight.

We simulated the GEM problem using the collisionless adiabatic
ten-moment pair plasma model supplemented with a globally
prescribed rate of pressure tensor isotropization.
We have neglected all diffusivities and relaxation terms
except isotropization.
We varied the rate of pressure isotropization and
studied the resulting variation in the rate of reconnection
and the contributions of the terms in Ohm's law.

\section{Model}

\def\Div{\nabla\cdot}
\def\curl{\nabla\times}
\def\Energy{\mathbb{E}}
\def\Pressure{\mathbb{P}}
\def\Sym{\,\mathrm{Sym}\,}
\def\u{\mathbf{u}}
\def\R{\mathbf{R}}
\def\RR{\mathbb{R}}
\def\B{\mathbf{B}}
\def\E{\mathbf{E}}
\def\J{\mathbf{J}}
\def\Q{\mathbb{Q}}
Generic physical equations for the ten-moment two-fluid model are:
\begin{itemize}
\item conservation of mass for each species:
\begin{alignat*}{5}
   &\partial_t \rho_s + \Div(\rho_s\u_s) &&= 0,
\end{alignat*}
\item conservation of momentum for each species:
\begin{alignat*}{5}
   &\partial_t (\rho_s\u_s) + \Div(\rho_s\u_s\otimes\u_s + \Pressure_s)
     &&= {q_s\over m_s}\rho_s (\E + \u_s\times\B) + \R_s,
\end{alignat*}
\item evolution of the pressure tensor for each
species:\footnote{For conservation and shock-capturing purposes
we actually evolve the \emph{energy} tensor
$\Energy_s := \Pressure_s + \rho_s\u_s\u_s$ rather than
directly evolving the pressure tensor.}
\begin{alignat*}{5}
   &\partial_t \Pressure_s 
     + \Div(\u_s\Pressure_s) + 2\Sym(\Pressure_s\cdot\nabla\u_s)
     + \Div\Q_s
     &&= 2\Sym({q_s\over m_s}\Pressure_s\times\B) + \RR_s,
\end{alignat*}
\item Maxwell's equations for evolution of electromagnetic field:
\begin{alignat*}{5}
   &\partial_t \B + \curl\E = 0,
\\ &\partial_t \E - c^2\curl\B = -\J/\epsilon,
\end{alignat*}
\item and Maxwell's divergence constraints:
\begin{alignat*}{5}
   &\Div\B &&= 0,
\\ &\Div\E &&= \sigma/\epsilon.
\end{alignat*}
\end{itemize}
In these equations,
$\Sym$ denotes the symmetric part of the argument tensor
(i.e.\ the average over all permutations of subscripts),
$\vee$ denotes symmetric outer product
(i.e.\ the symmetric part of the tensor product),
and
$i$ and $e$ are positive (``ion'') and negative (``electron'') species indices;
for species $s\in\{i,e\}$,
$q_s=\pm e$ is particle charge,
$m_s$ is particle mass,
$n_s$ is particle number density,
$\rho_s = m_s n_s$ is mass density,
$\sigma_s = q_s n_s$ is charge density,
$\u_s\rho_s$ is momentum,
$\J_s = \u_s\sigma_s$ is current density, and
$\Pressure_s$ is a pressure tensor;
$\B$ is magnetic field,
$\E$ is electric field,
$c$ is the speed of light,
$\epsilon$ is vacuum permittivity,
$\J = \J_i + \J_e$ is net current density, and
$\sigma = \sigma_i + \sigma_e$ is net charge density.
To close the system, constitutive relations must be supplied for
the generalized heat fluxes
$\Q_s$,
the interspecies drag force on the ions
$\R_i=-\R_e$,
and $\RR_s$, the production of generalized thermal energy
due to collisions.
We nondimensionalize these equations,
choosing the timescale to be the gyroperiod of a typical particle of mass $1$, and
choosing the typical velocity to be a typical Alfv\'en speed.  
The nondimensionalized equations retain the form of the dimensional equations
above, with the simplifications that $e=1$, $m_i+m_e=1$, and
${1\over \epsilon}=c^2$.

In our closure we assume that $\R_s = 0$, and to provide
for isotropization we let
\def\tr{\mathrm{tr}\,}
\def\idtens{\mathbb{I}}
\begin{gather*}
  \RR_s = \frac{1}{\tau_s}\bigg({1 \over 3}(\tr\Pressure_s) \idtens - \Pressure_s\bigg),
\end{gather*}
where $\tau_s$ is the isotropization period of species $s$,
$\tr$ denotes tensor trace, and
$\idtens$ is the identity tensor.
In our present work we assume that 
$\Q_s=0$.  This assumption might not be satisfactory, though:
the particle simulations of Hesse et al. \cite{article:HeKuBi04} showed,
at least for the case of guide-field electron-proton
reconnection, that generalized heat flux contributions
to the evolution of the pressure tensor
are necessary to obtain an appropriate approximation
for the pressure nongyrotropy near the X-point.
We therefore plan to investigate C. David Levermore's 
closure \cite{Le09},
\begin{gather*}
  \Q_s = \frac{9}{5} (\nu_0-\nu_1) \idtens\vee\tr\big(\nabla\vee\Theta_s^{-1}\big)
       + 3 \nu_1\Big(\nabla\vee \Theta_s^{-1}\Big),
\end{gather*}
where $\Theta_s:=\Pressure_s/\rho_s$ and $\nu_0\simeq\nu_1$ is proportional to collision frequency.  We expect to determine whether we can
get fast reconnection without isotropization by using such
a non-vanishing generalized heat flux.

\subsection{Ohm's law.}

Combining the momentum equations gives
net current balance.  Assuming quasineutrality (zero net charge)
and solving for electric field gives Ohm's law for the electric field:
\def\mti{\tilde m_i}
\def\mte{\tilde m_e}
\begin{alignat*}{4}
   \E = &\frac{\mti+\mte}{\rho}(-\R_i)
           &&\text{ (resistive term)}
    \\  &+ \B\times\u
           &&\text{ (ideal term)}
    \\  &+ \frac{\mti-\mte}{\rho}\J\times\B
           &&\text{ (Hall term)}
  \\ &+ \frac{1}{\rho}\Div( \mte\Pressure_i - \mti\Pressure_e)
           && \text{ (pressure term)}
    \\ &+ \frac{\mti\mte}{\rho}
         \Big(\partial_{t} \J +\Div\big(
            \u\J + \J\u + \frac{\mte-\mti}{\rho}\J\J\big)\Big)
        &&  \text{ (inertial term)},
\end{alignat*}
where
$\mti:={m_i\over e}$ and $\mte:={m_e\over e}$
and the resistive term is usually assumed to be of
the form $\eta\cdot\J$, i.e., a linear function of current.

\section{GEM magnetic reconnection challenge problem}

The GEM magnetic reconnection challenge problem studies the
evolution of 2-dimensional plasma in a rectangular box aligned
with the coordinate axes and centered at the origin.
The top and bottom of the box are conducting walls and periodic
symmetry in the $x$-axis defines the width of the box. The plasma
is initially in near-equilibrium. The upper half of the box is
occupied by strong magnetic field lines pointing to the right
and the lower half is occupied by strong magnetic field
lines pointing to the left, separated by a thin, potentially volatile
transition layer along the $x$-axis. Some studies
(e.g.\ \cite{article:ChSiLuZo08}) add a constant
out-of-plane component to the magnetic field, called a ``guide
field''. We do not have a guide field (nor did the original GEM
problem or the studies we are trying to replicate).


\subsection{Domain.}
The computational domain is the rectangular domain
$[-L_x/2,L_x/2]\times[-L_y/2,L_y/2]$.
The problem is symmetric under 180 degree rotation around the origin,
and in the case of zero guide field is also symmetric under
reflection across either the horizontal or vertical axis.
In the original GEM problem, $L_x=8\pi$ and $L_y=4\pi$.
We halved the dimensions, so that $L_x=4\pi$ and $L_y=2\pi$.

\subsection{Boundary conditions.}
The domain is periodic along the $x$-axis.
The boundaries parallel to the $x$-axis
are thermally insulating conducting wall boundaries.  
A conducting wall boundary is a solid wall boundary
(with slip boundary conditions in the case of ideal plasma)
for the fluid variables, and the electric field at
the boundary has no component parallel to the boundary.
We also assume that magnetic field does not penetrate the boundary.

\subsection{Model Parameters.}
We set the speed of light to 10
(rather than 20 as in \cite{article:BeBh05})
and set the mass of each species to 0.5
(rather than the GEM values of $1$ for ions
and $1/25$ for electrons).

\subsection{Initial conditions.}
The initial conditions are a perturbed Harris sheet equilibrium.
The unperturbed equilibrium is given by
\def\e{\mathbf{e}}
\def\E{\mathbf{E}}
\def\J{\mathbf{J}}
\def\sech{\,\mathrm{sech}}
\begin{align*}
    \B(y) & =B_0\tanh(y/\lambda)\e_x,
  & p(y) &= \frac{B_0^2}{2 n_0} n(y),
 \\ n(y) &= n_i(y) = n_e(y)
            = n_0(1/5+\sech^2(y/\lambda)),
  & p_e(y) &= \frac{T_e}{T_i+T_e}p(y),
 \\ \E & =0,
  & p_i(y) &= \frac{T_i}{T_i+T_e}p(y).
\end{align*}
On top of this the magnetic field is perturbed by
\begin{gather*}
   \delta\B =-\e_z\times\nabla(\psi),\ \ \hbox{ where } \ \ 
   \psi(x,y)=\psi_0 \cos(2\pi x/L_x) \cos(\pi y/L_y).
\end{gather*}
In the GEM problem the initial condition constants are
\begin{align*}
    T_i/T_e &= 5,
  & \lambda&=0.5,
  & B_0&=1,
  & n_0&=1,
  & \psi_0&=B_0/10;
\end{align*}
we reset the initial temperature ratio to 1 to get symmetry between
the species, and we set $\psi_0$ to $r_s^2 B_0/10$, where
$r_s=0.5$ is our domain rescaling factor, so that in the vicinity
of the X-point our initial conditions agree (up to first-order
Taylor expansion) with the initial conditions of the GEM problem.

\section{Properties of the GEM problem}

\subsection{Reconnected flux.}

We define magnetic reconnection to be the loss
of magnetic flux through the vertical axis
into the first quadrant.
Using Faraday's law, $\partial_t \B + \curl\E=0$,
one can show that the rate of reconnection is minus the value of the out-of-plane
component of the electric field at the origin (i.e.\ the X-point) \cite{article:JoRo08a}.

\subsection{Ohm's law at the origin.}

Since the electric field at the origin is the rate of reconnection,
we are lead to study the terms of Ohm's law for the electric field
at the origin.  Since the problem is symmetric under 180 degree rotational symmetry about the
origin, only the out-of-plane component of vectors is nonzero at the origin.
In Ohm's law only the out-of-plane components of the resistive, pressure, and inertial terms survive.

As a proxy for Ohm's law\footnote{
Ohm's law involves the approximating assumption of
quasineutrality, whereas the momentum equation
holds exactly and the inertial term reduces to a simpler form
at the origin.},
we select a species and write the momentum equation
solved for the electric field:
\def\origin{\mathrm{origin}}
\begin{gather*}
d_t(\mathrm{reconnected\ flux}) = \E_3\big|_\origin
   = \bigg[\frac{-\R_i}{e n_i} + \frac{\Div\Pressure_i}{e n_i}
  + \frac{m_i}{e} \partial_t\u_i\bigg]_3\Bigg|_\origin.
\end{gather*}
In a perfectly collisionless, gyrotropic plasma, the resistive
term and pressure divergence vanish, and
\emph{reconnected flux should exactly track with species velocity}
(a proxy for the current) at the origin.

\section{Results}

We simulated the GEM magnetic reconnection challenge problem
for a pair plasma, varying the mesh resolution and varying
the rate of isotropization from zero to instantaneous. We plotted
the contribution of proxy Ohm's law terms to the reconnected flux.
We find that for a broad intermediate range of isotropization rates
reconnection is fast and that the pressure term makes the dominant
contribution to reconnected flux.

When isotropization is very slow or absent, there is oscillatory
exchange between the inertial term and the pressure term at roughly a
typical gyrofrequency, and reconnection proceeds at a slow to moderate
rate (our simulations for the case of no isotropization are
not sufficiently resolved).
For a broad intermediate range of isotropization, there is little
oscillation and, 
in agreement with PIC simulations (see \cite{article:BeBh05, article:BeBh07}),
the pressure divergence dominates and provides for faster reconnection.
As the rate of isotropization becomes very fast, however, the
pressure divergence is forced to vanish and
the inertial term is the only remaining term in the
equation that can provide for reconnected flux.
This forces the current at the origin to ramp up in
track with reconnected flux.
The system seems unable to sustain
this ramp-up in current, however, and numerical instability 
kicks in, as evidenced by the sudden appearance of
strong, very rapid oscillatory exchange (less evident in
the accumulation integrals shown) between
the inertial term and the residual. 
The numerical residual displaces the inertial term,
allowing the the current to peak and then decay while
the reconnected flux maintains the same smooth, rapid ascent
seen for intermediate isotropization rates, as if
unconcerned whether pressure, inertia, numerical
resistivity -- or even anomalous resistivity, as our
cursory investigations suggest -- provides for its determined course.
We can conclude that fast reconnection at least commences
in an isotropic pair plasma model, and we conjecture that
the numerical instability we see corresponds to some physical
(perhaps streaming?) instability that provides for
an effective anomalous resistivity (whose functional form
we have not analyzed). 
As expected, the five-moment simulations show agreement with
instantaneous relaxation of the ten-moment system to isotropy.

\begin{figure}[htbp!]
  \includegraphics[width=3.25in, height=1.6in]{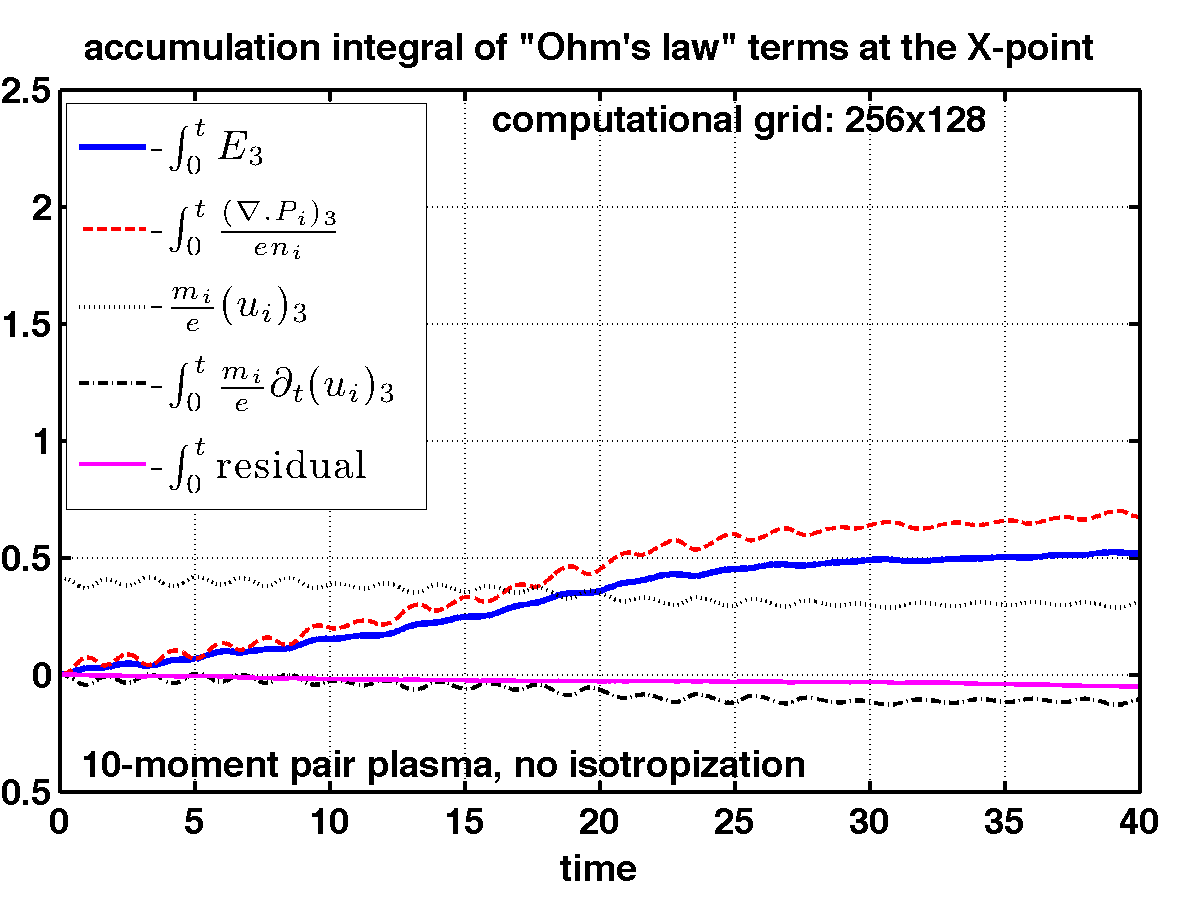}
  \includegraphics[width=3.25in, height=1.6in]{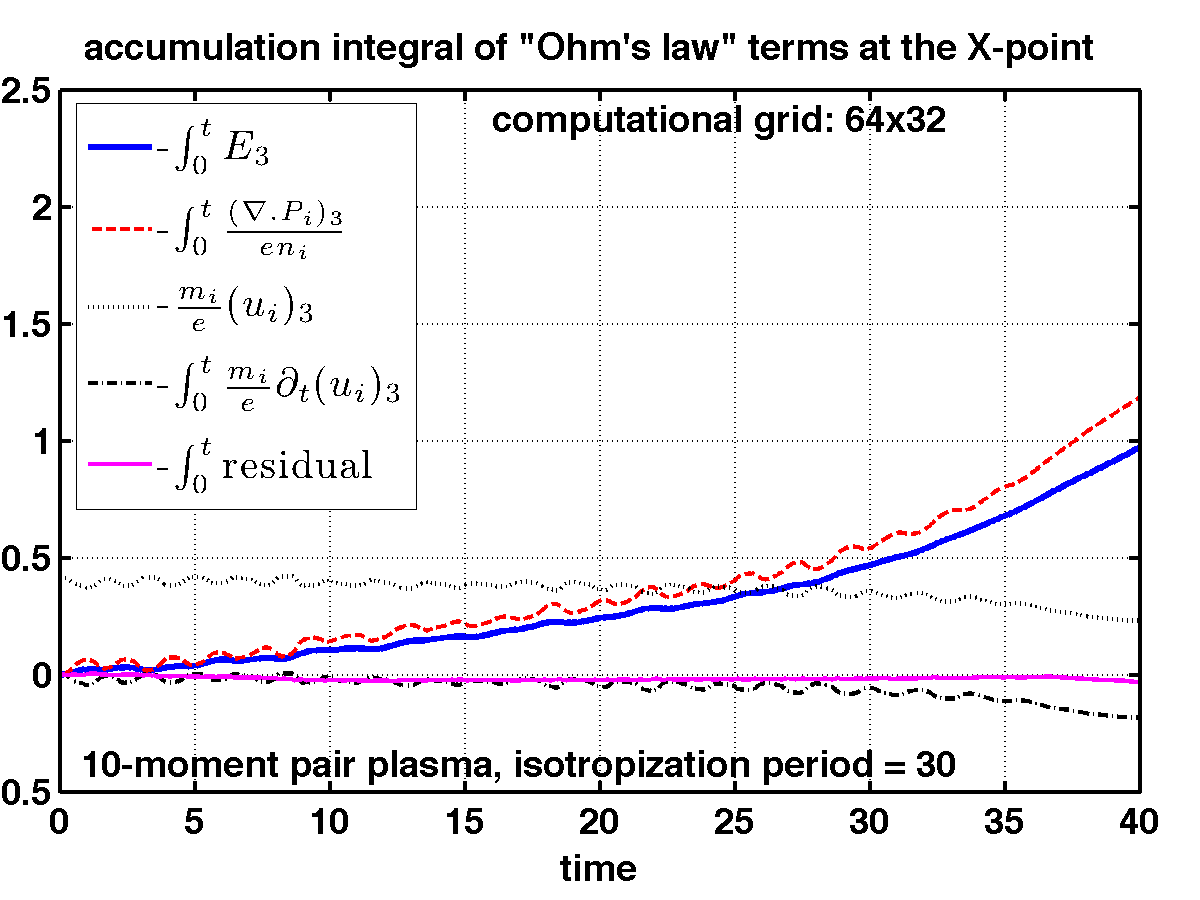}
  \includegraphics[width=3.25in, height=1.6in]{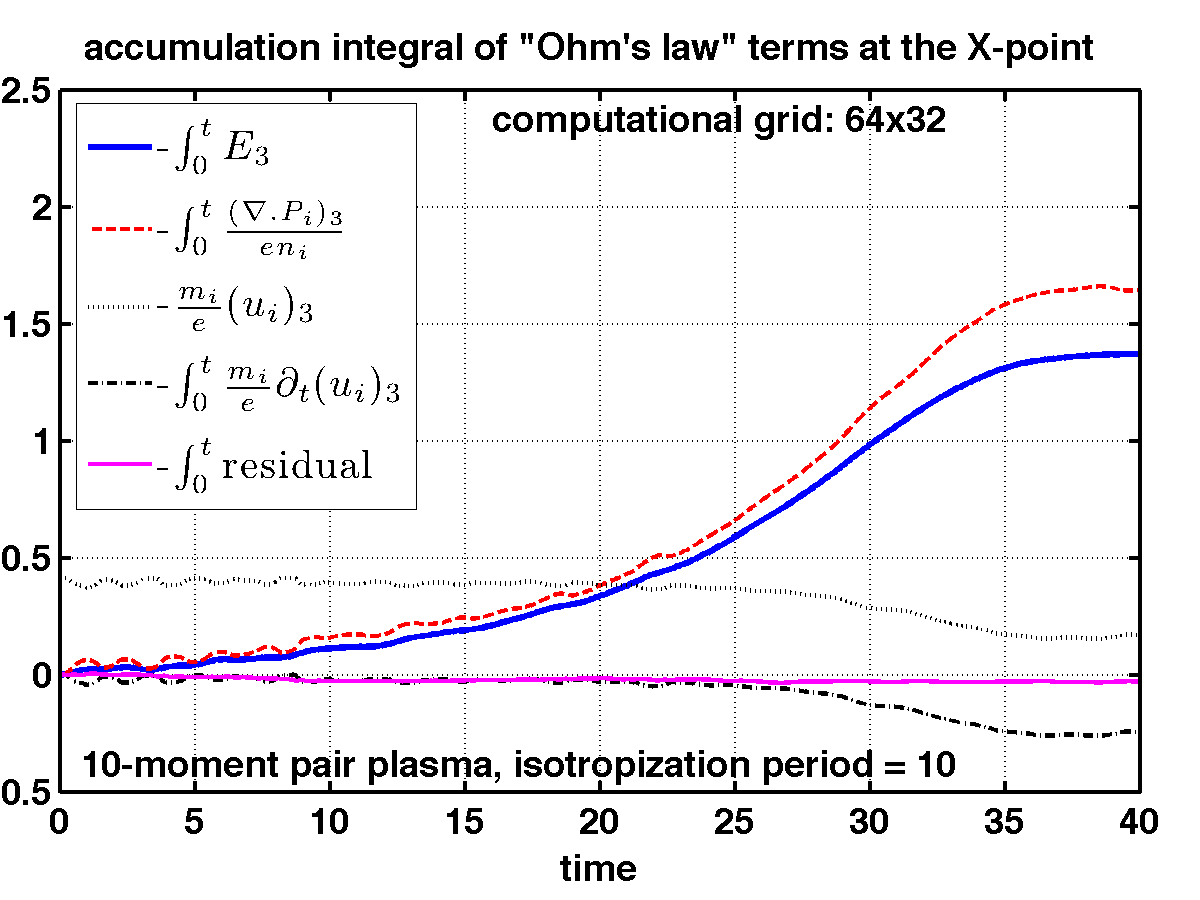}
  \includegraphics[width=3.25in, height=1.6in]{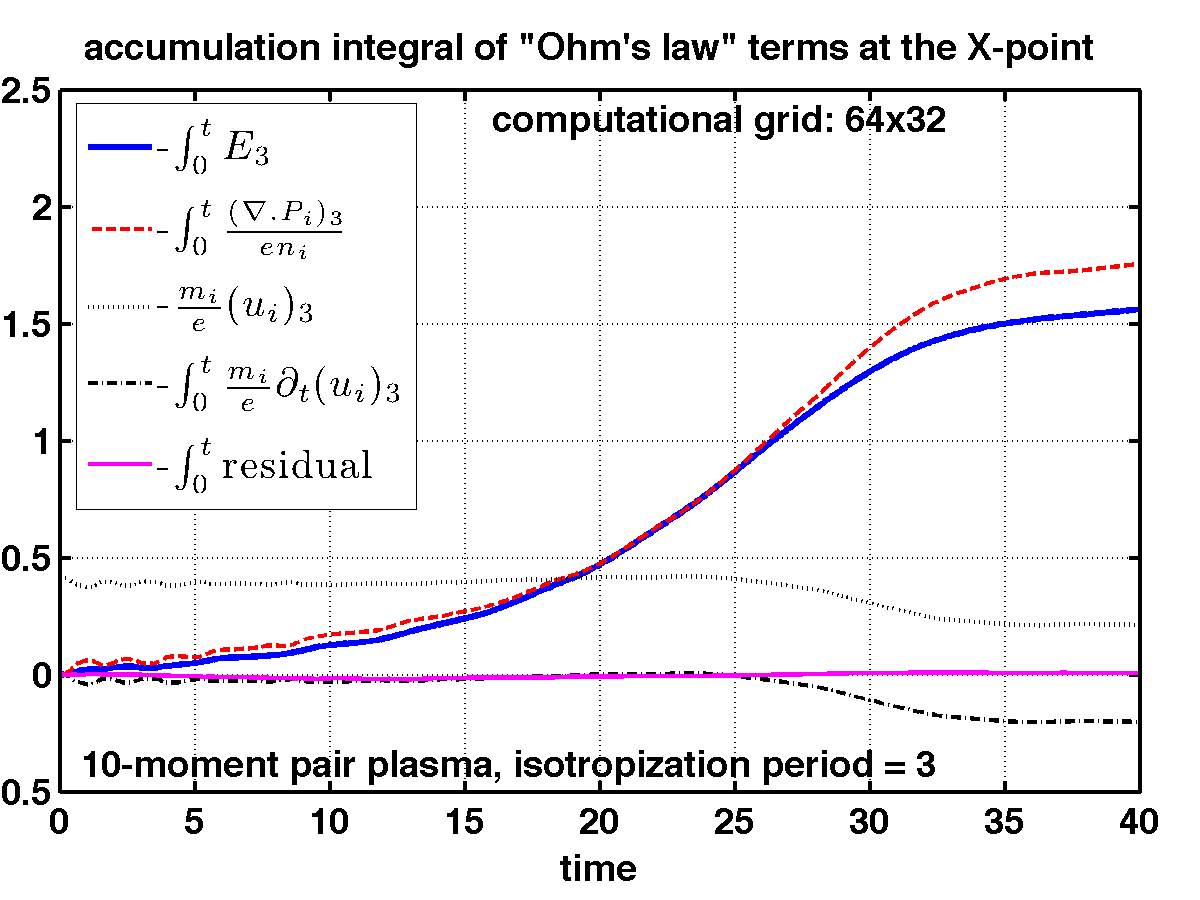}
  \includegraphics[width=3.25in, height=1.6in]{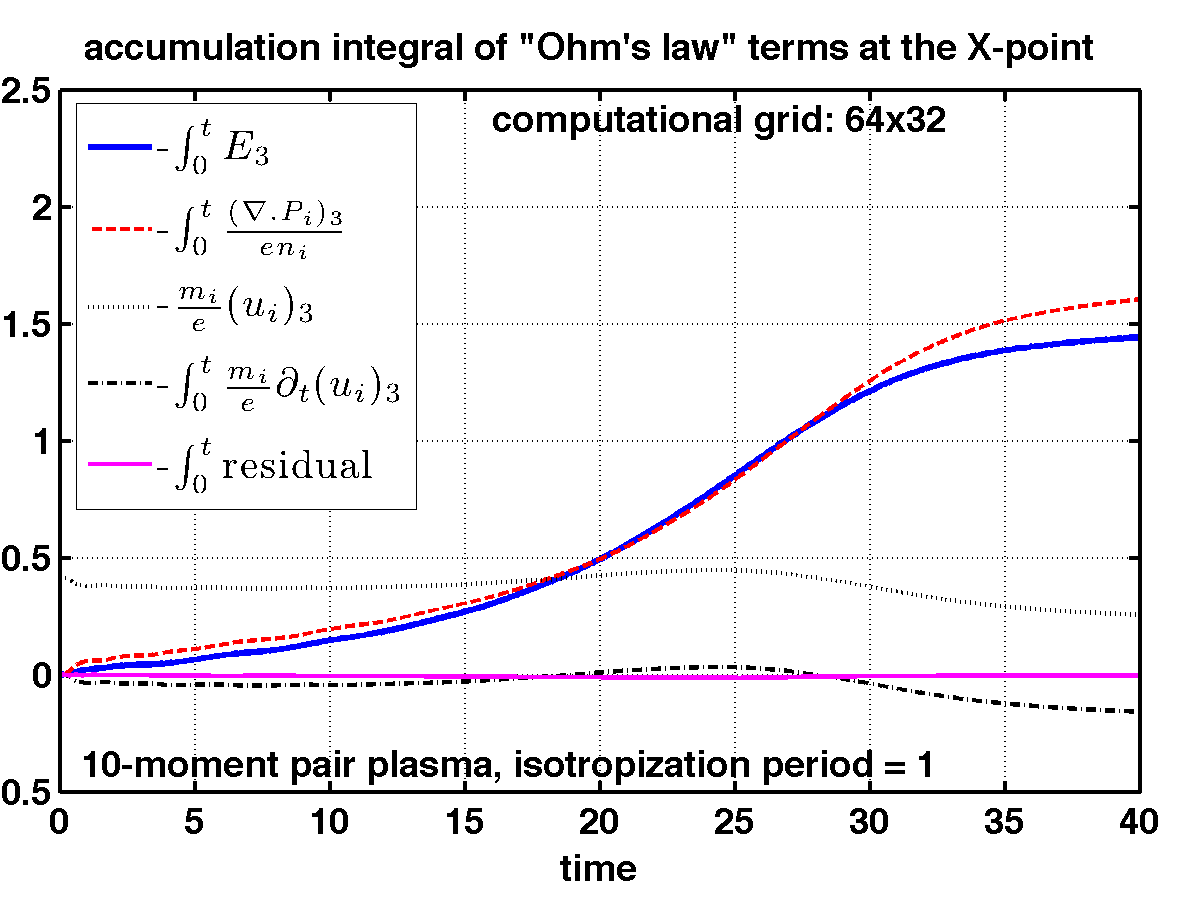}
  \includegraphics[width=3.25in, height=1.6in]{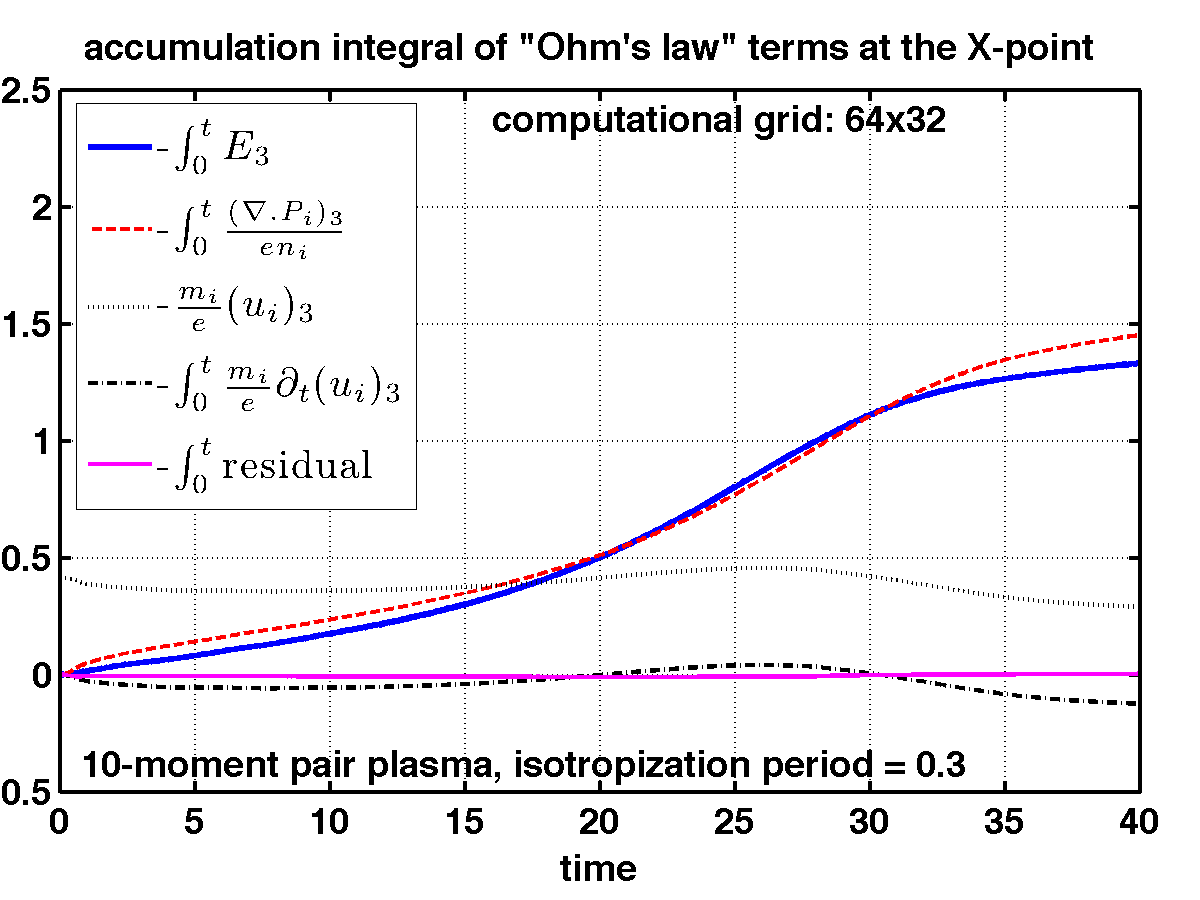}
  \includegraphics[width=3.25in, height=1.6in]{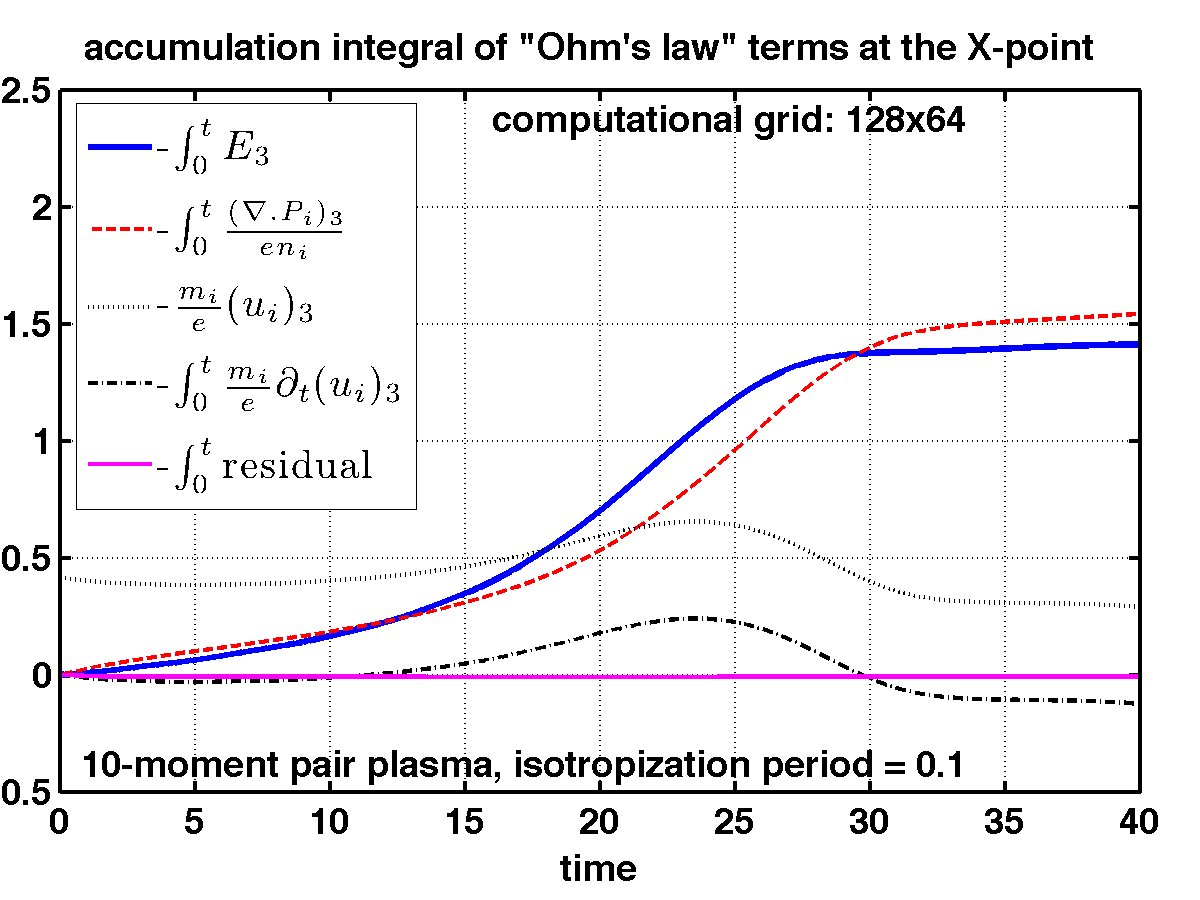}
  \includegraphics[width=3.25in, height=1.6in]{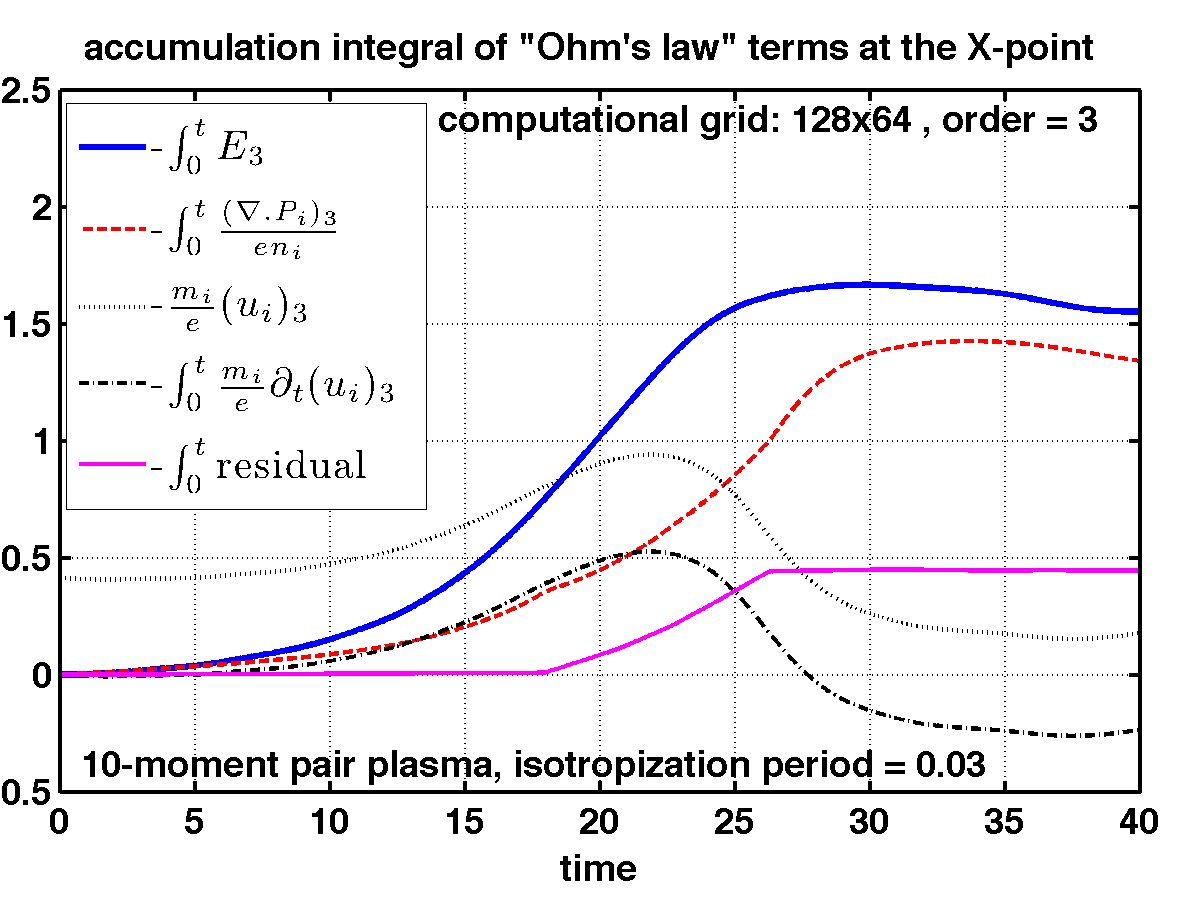}
  \includegraphics[width=3.25in, height=1.6in]{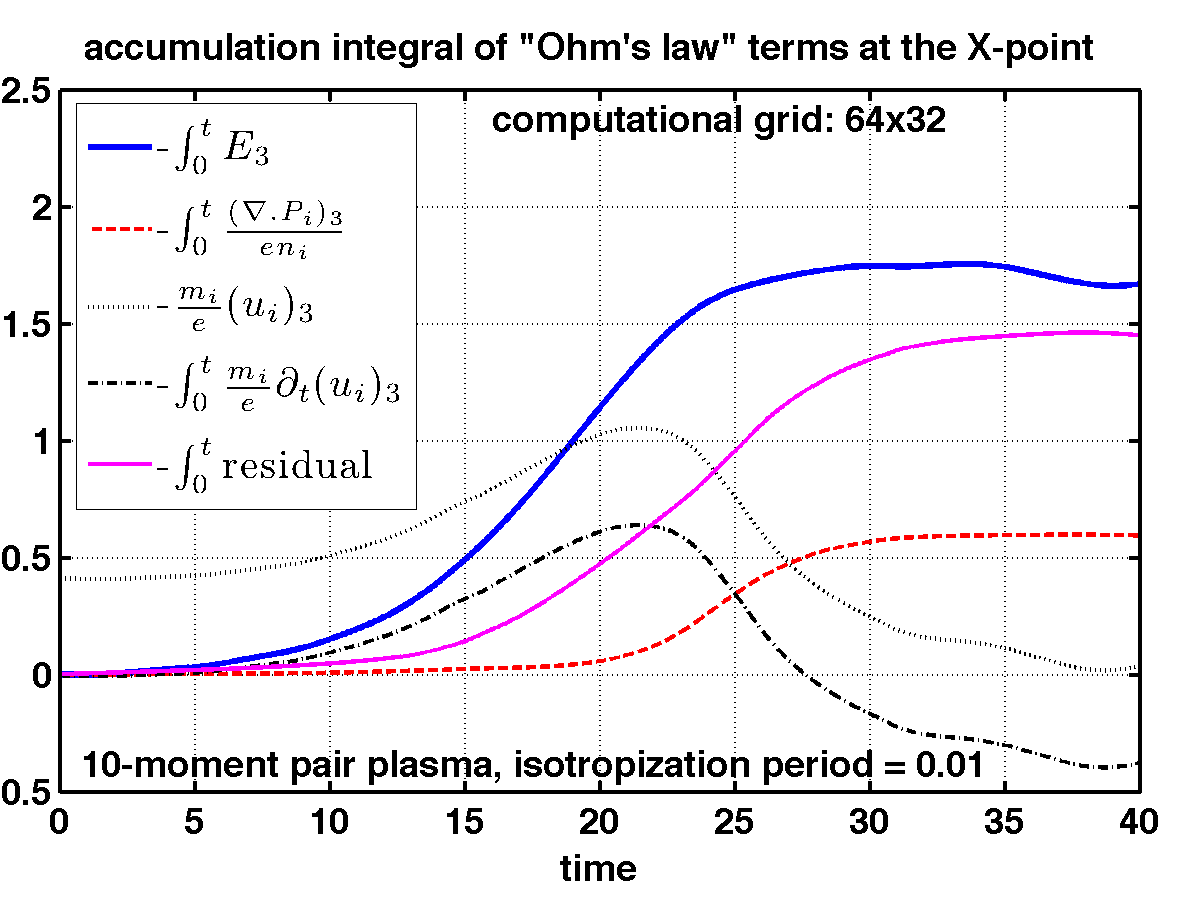}
  \includegraphics[width=3.25in, height=1.6in]{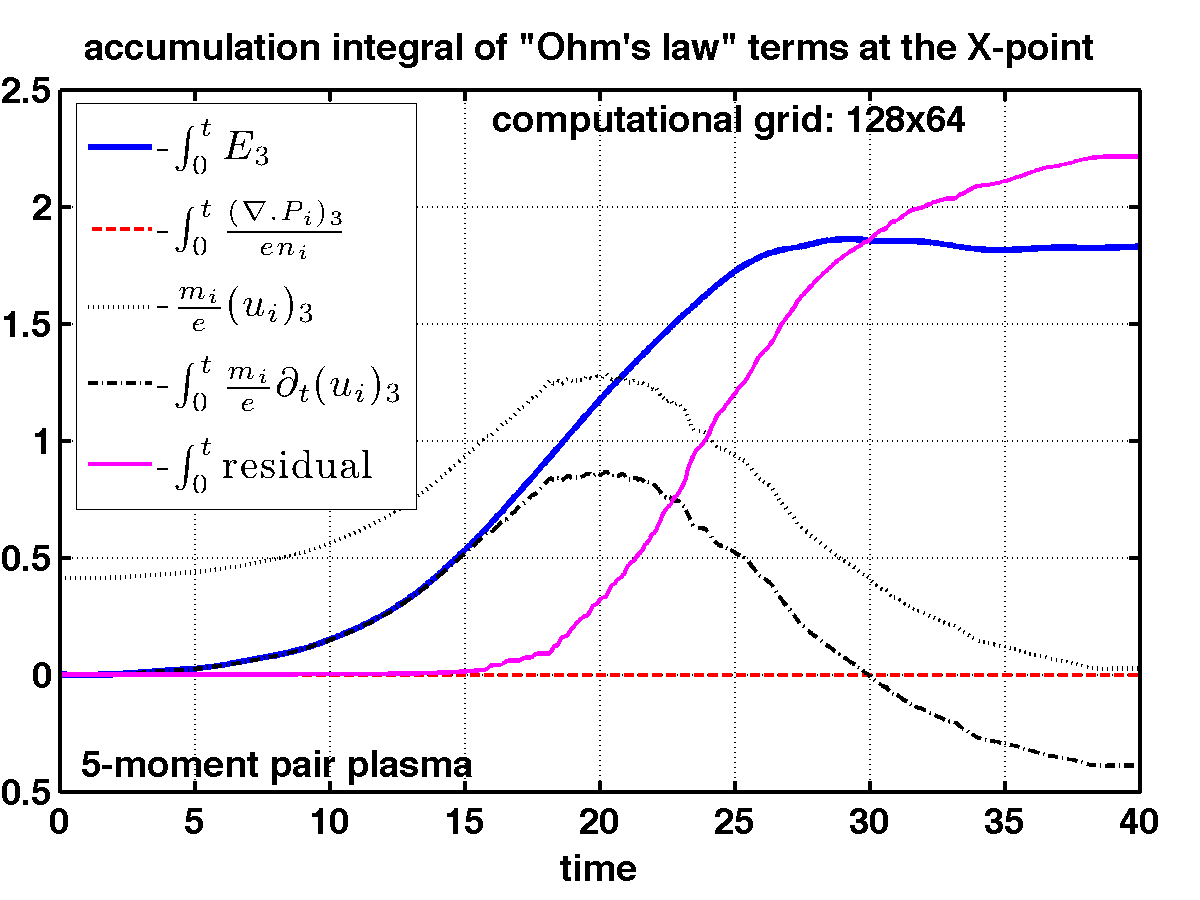}
  \label{f:10var_iso}
  \caption{\small Increase in reconnection rate as isotropization increases.
    For very slow or no isotropization the simulations we report are
    not sufficiently resolved. For instantaneous relaxation the
    positron bulk velocity ceases to track with reconnected
    flux and the residual becomes unacceptably large after
    $t=15$, indicating that we are no longer solving the momentum
    equation faithfully.}
  \vskip2pt \hrule
\end{figure}
\begin{figure}[htbp!]
  \includegraphics[width=3.25in, height=1.34in]{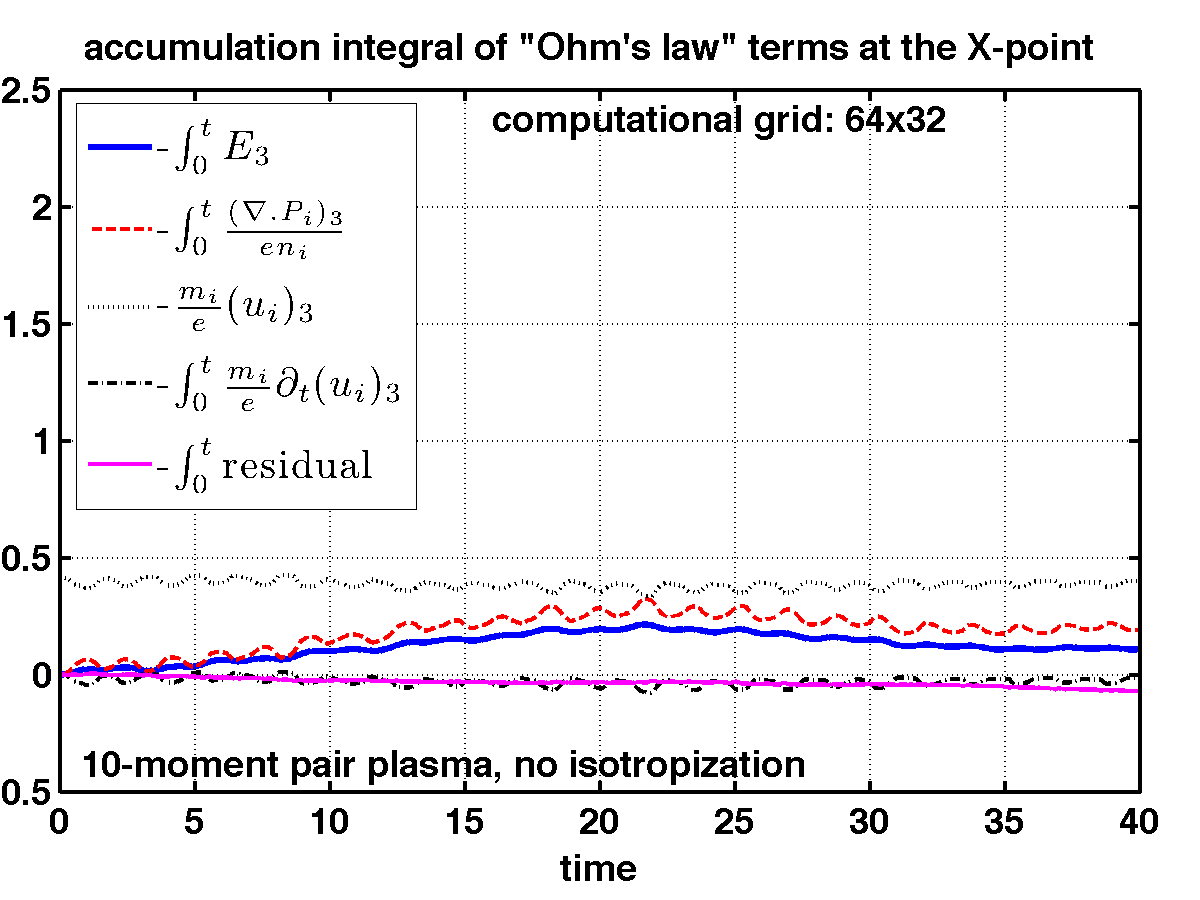}
  \includegraphics[width=3.25in, height=1.34in]{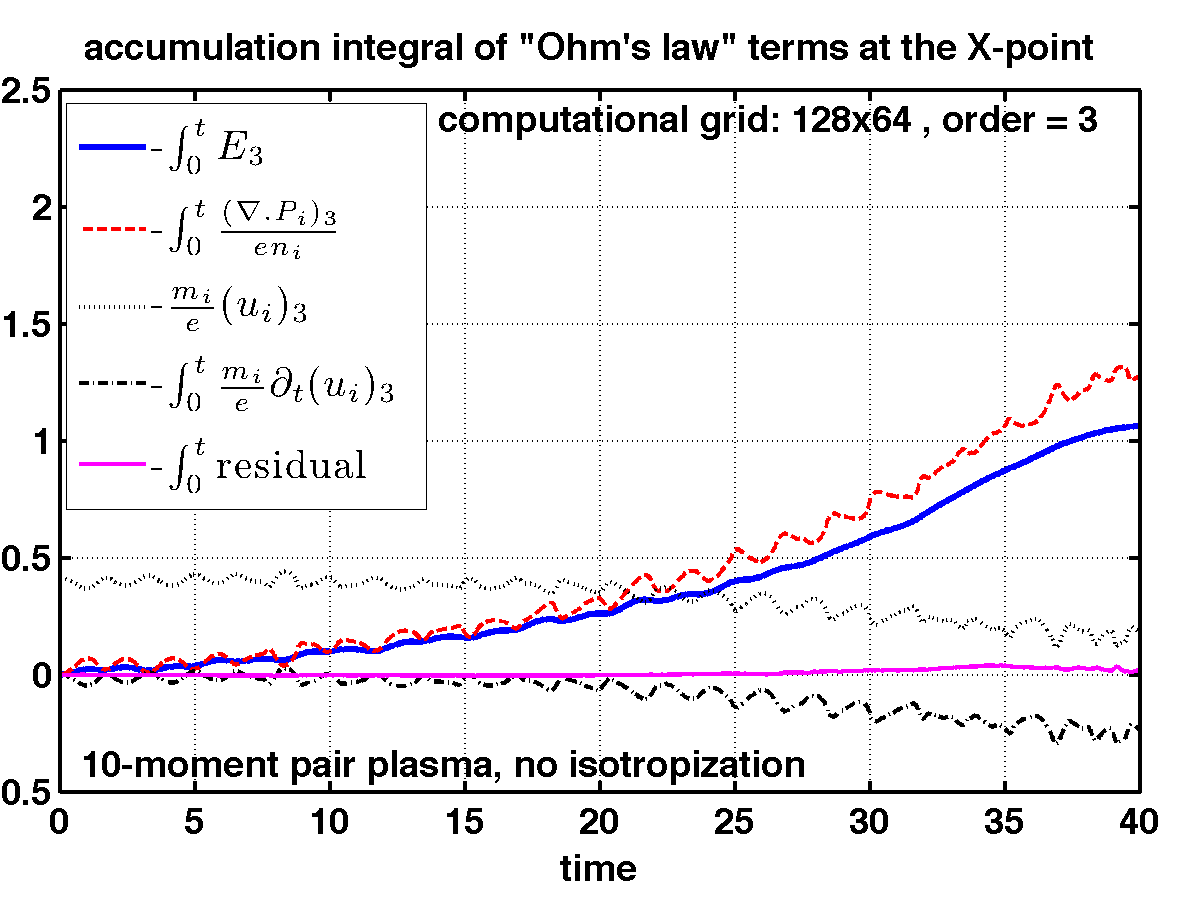}
  \includegraphics[width=3.25in, height=1.34in]{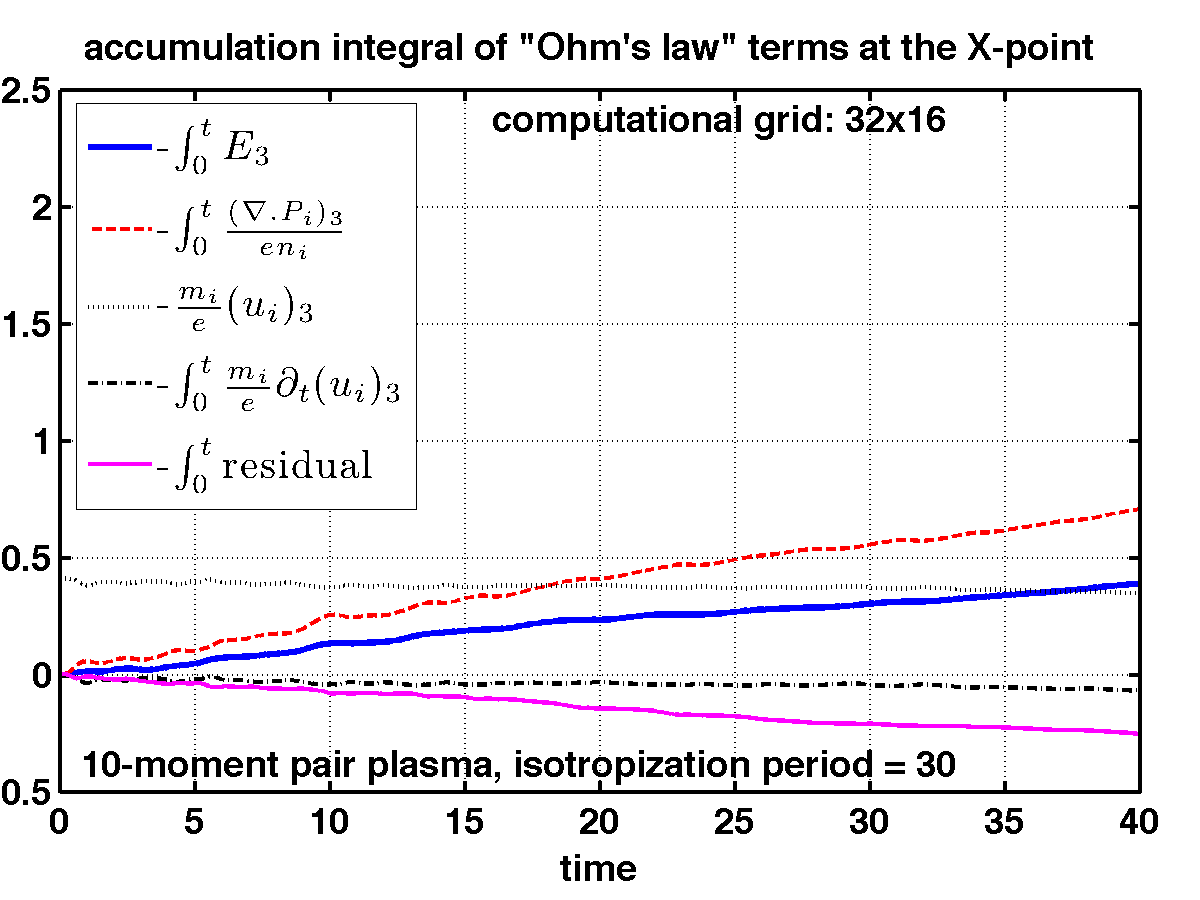}
  \includegraphics[width=3.25in, height=1.34in]{figures/accumint_p10_64x32_iso_period=30.png}
  \includegraphics[width=3.25in, height=1.34in]{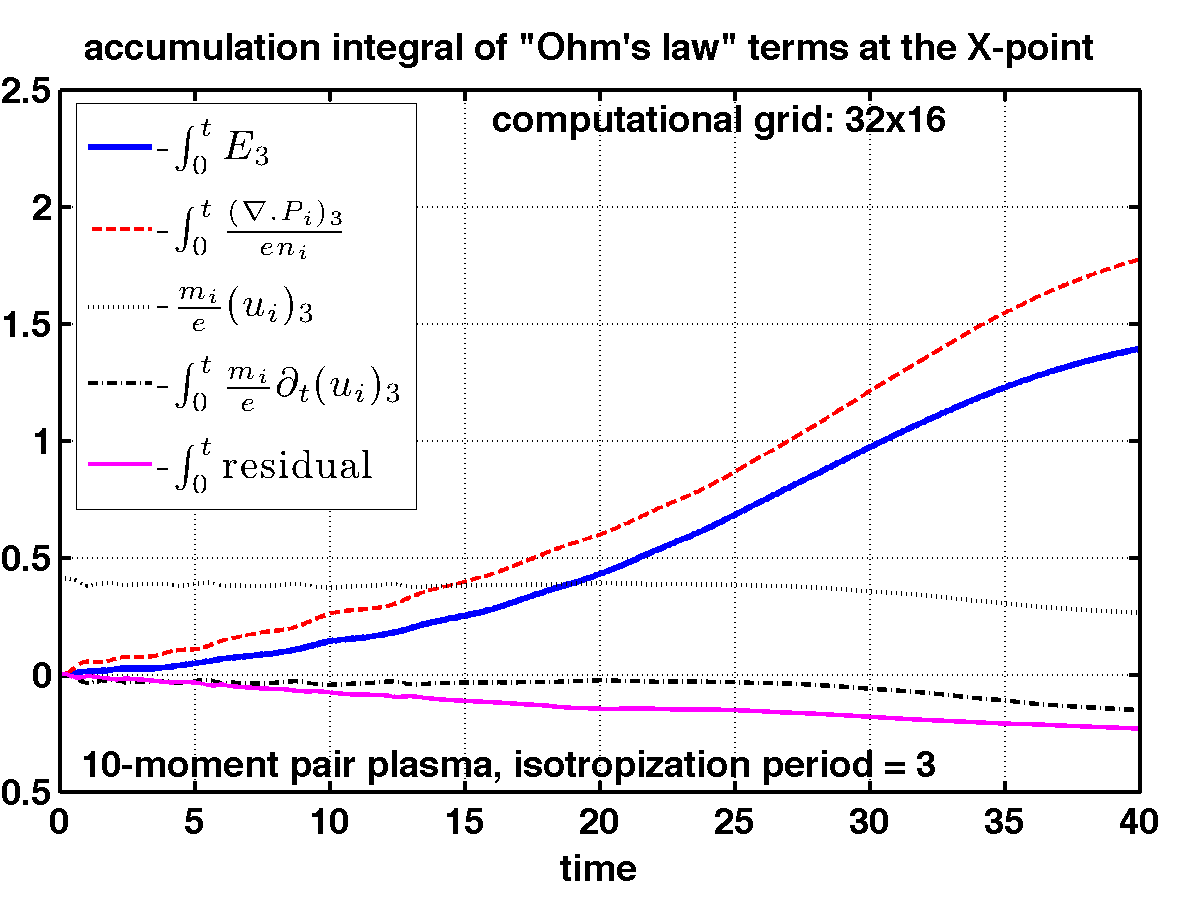}
  \includegraphics[width=3.25in, height=1.34in]{figures/accumint_p10_64x32_iso_period=3.png}
  \includegraphics[width=3.25in, height=1.34in]{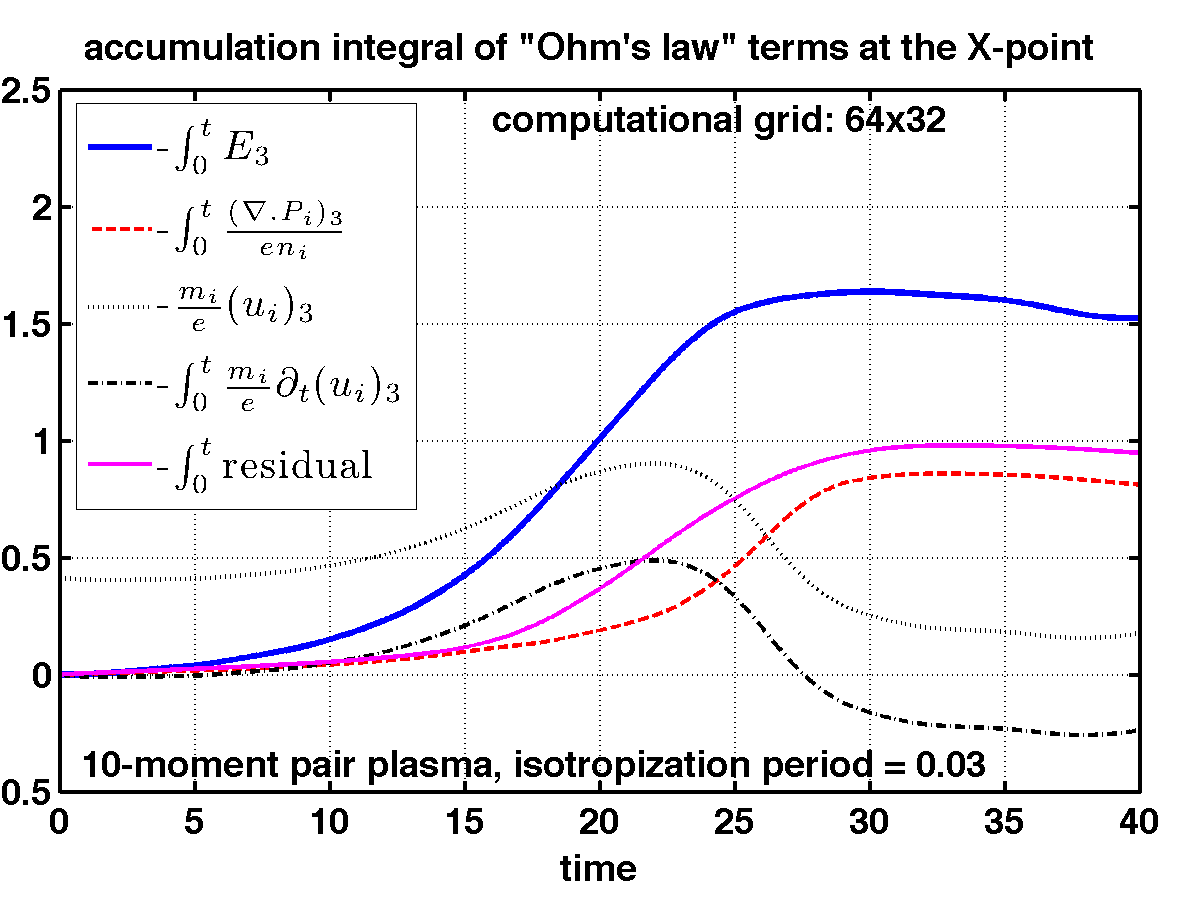}
  \includegraphics[width=3.25in, height=1.34in]{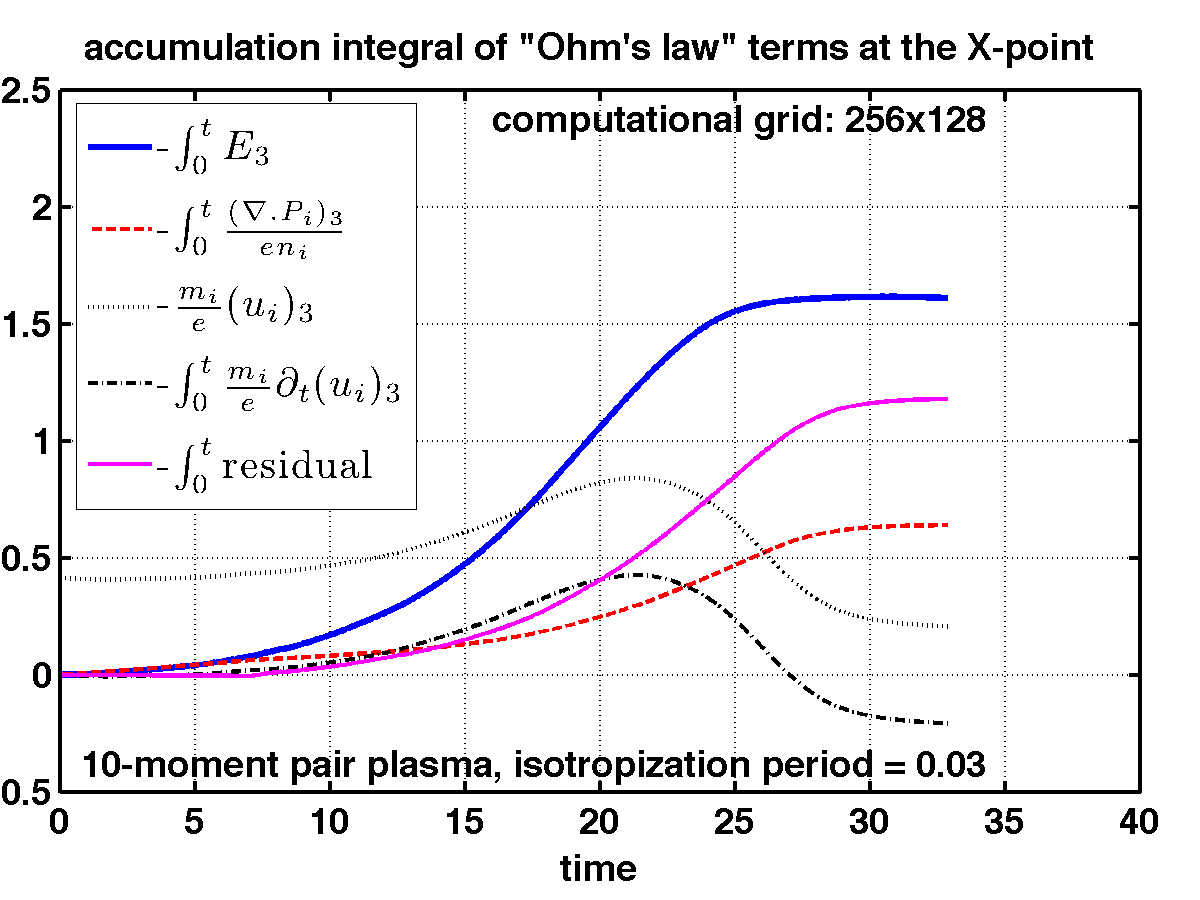}
  \includegraphics[width=3.25in, height=1.34in]{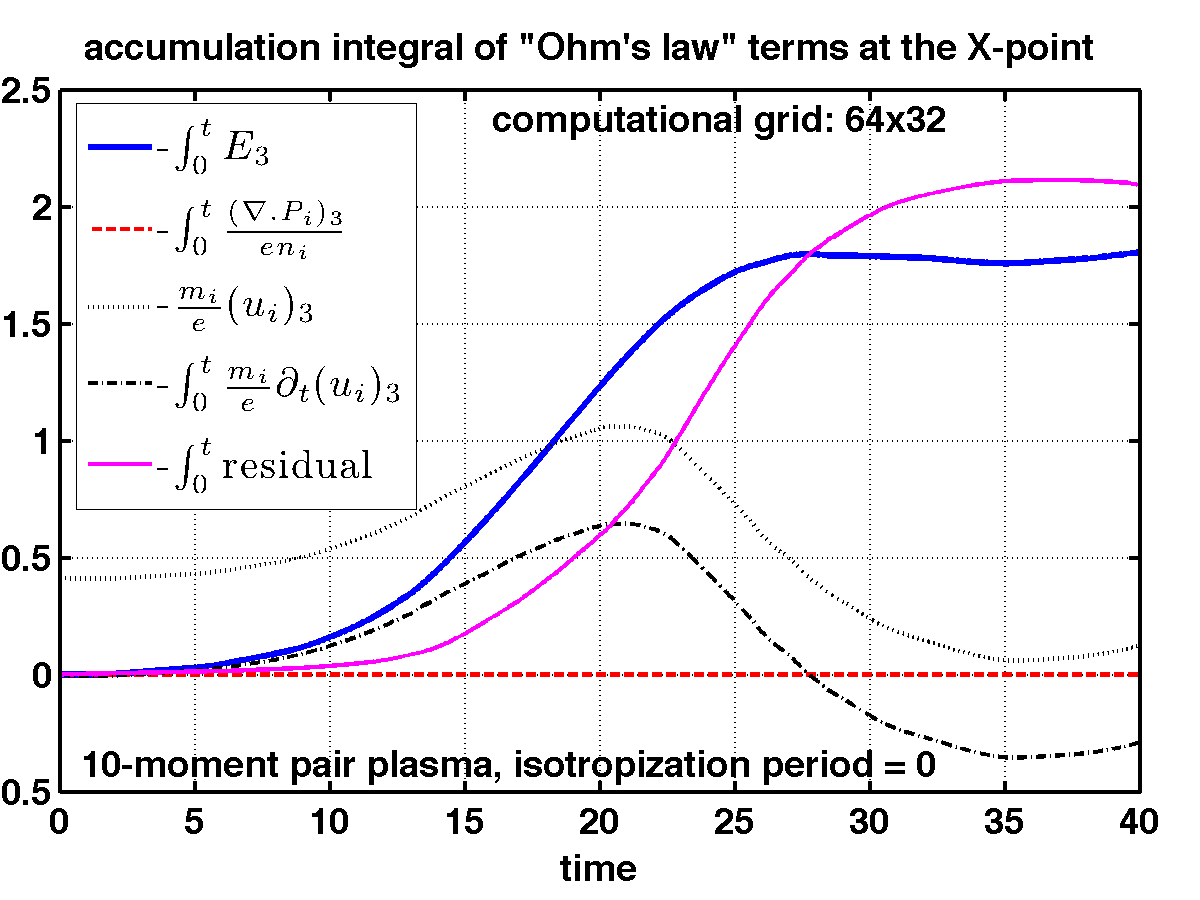}
  \includegraphics[width=3.25in, height=1.34in]{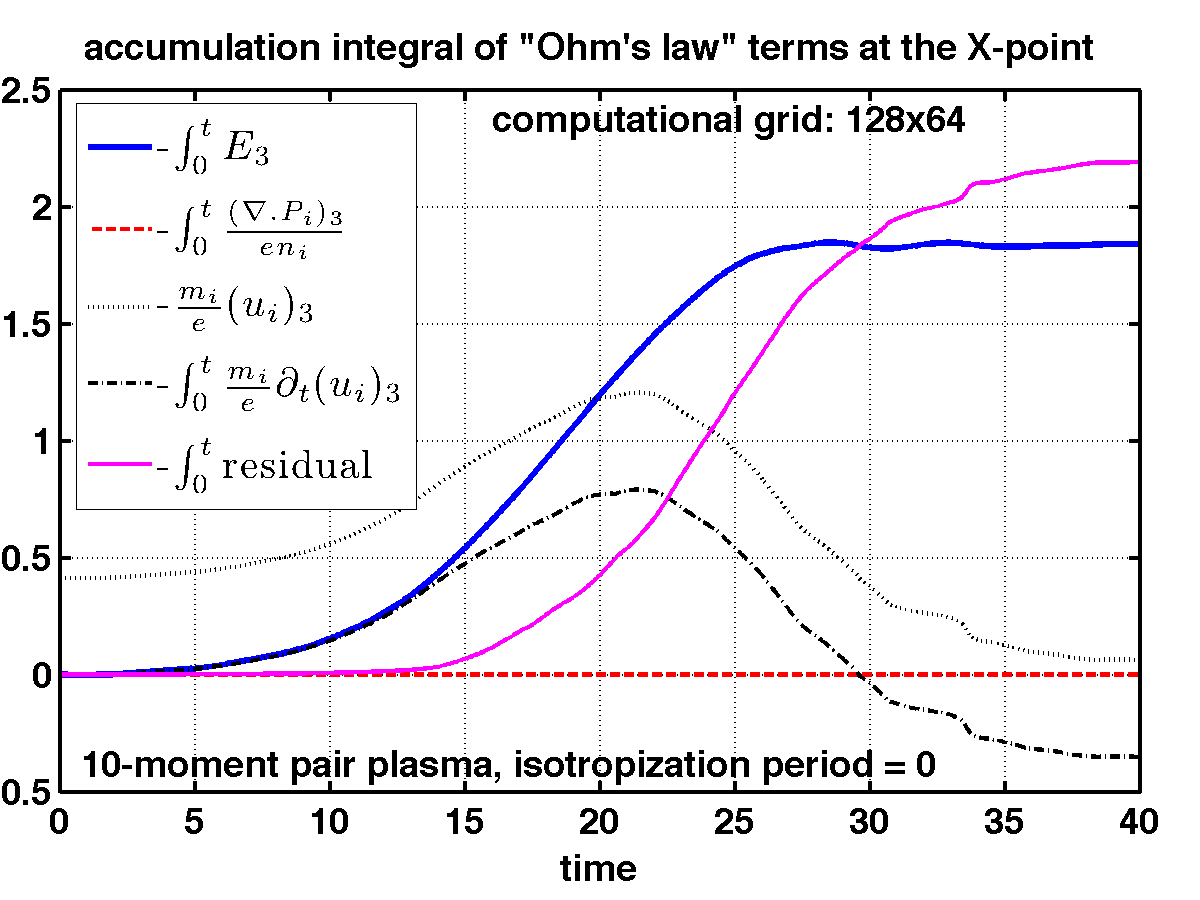}
  \includegraphics[width=3.25in, height=1.34in]{figures/accumint_p05_128x64.png}
  \includegraphics[width=3.25in, height=1.34in]{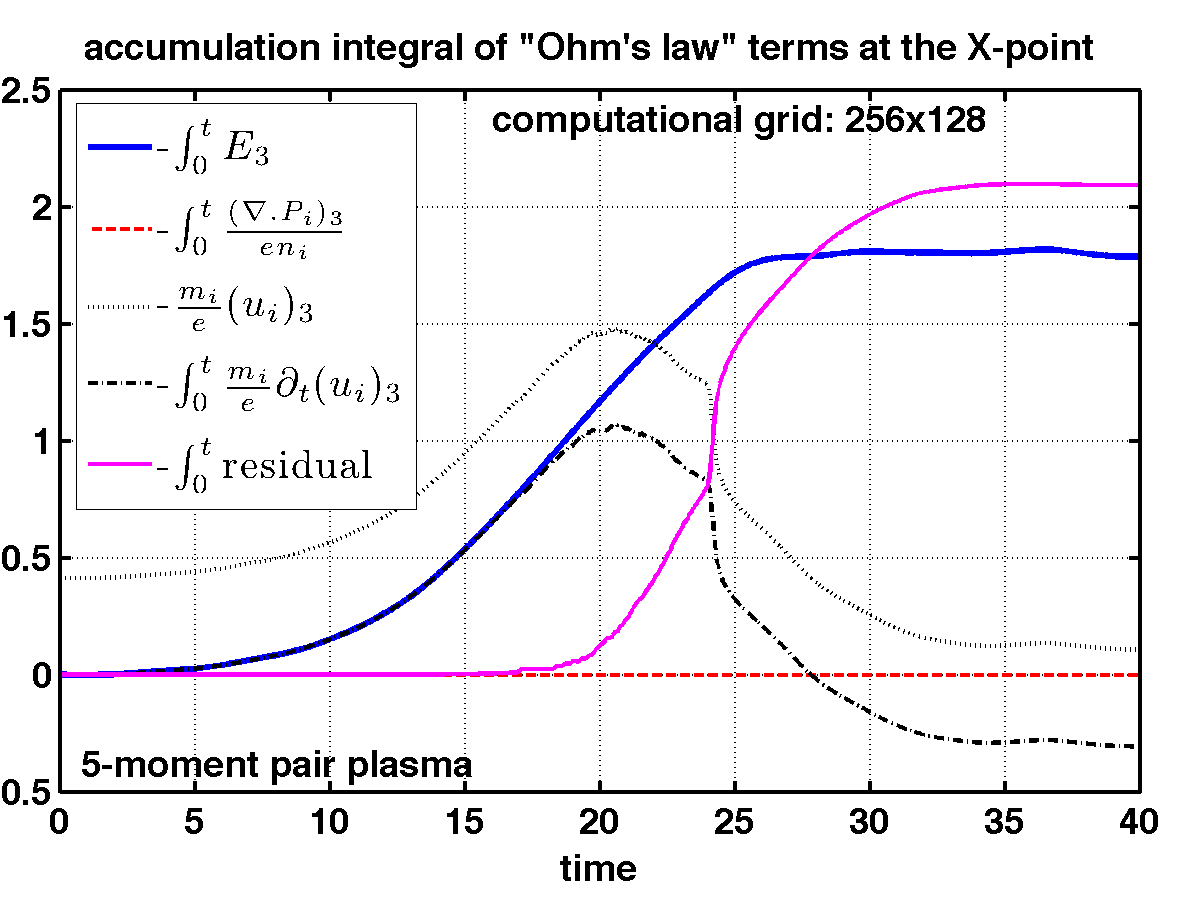}

  \caption{\small Convergence comparisons for coarse (left) and fine (right)
    meshes.  For intermediate isotropization rates a coarser mesh is
    satisfactory, but for no isotropization and for fast isotropization
    a finer mesh is required.  For 
    fast and instantaneous isotropization, even for a very fine mesh,
    the residual in ``Ohm's law'' (the momentum equation)
    becomes unacceptably large, overwhelming and displacing
    the inertial and pressure terms.
    The five-moment simulations
    and the instantaneously relaxed ten-moment simulations agree
    well where convergence is demonstrated.}
  \label{f:5vs10agreement}
  \vskip2pt \hrule
\end{figure}

\begin{figure}[htbp!]
  \includegraphics[width=3.25in, height=1.8in]{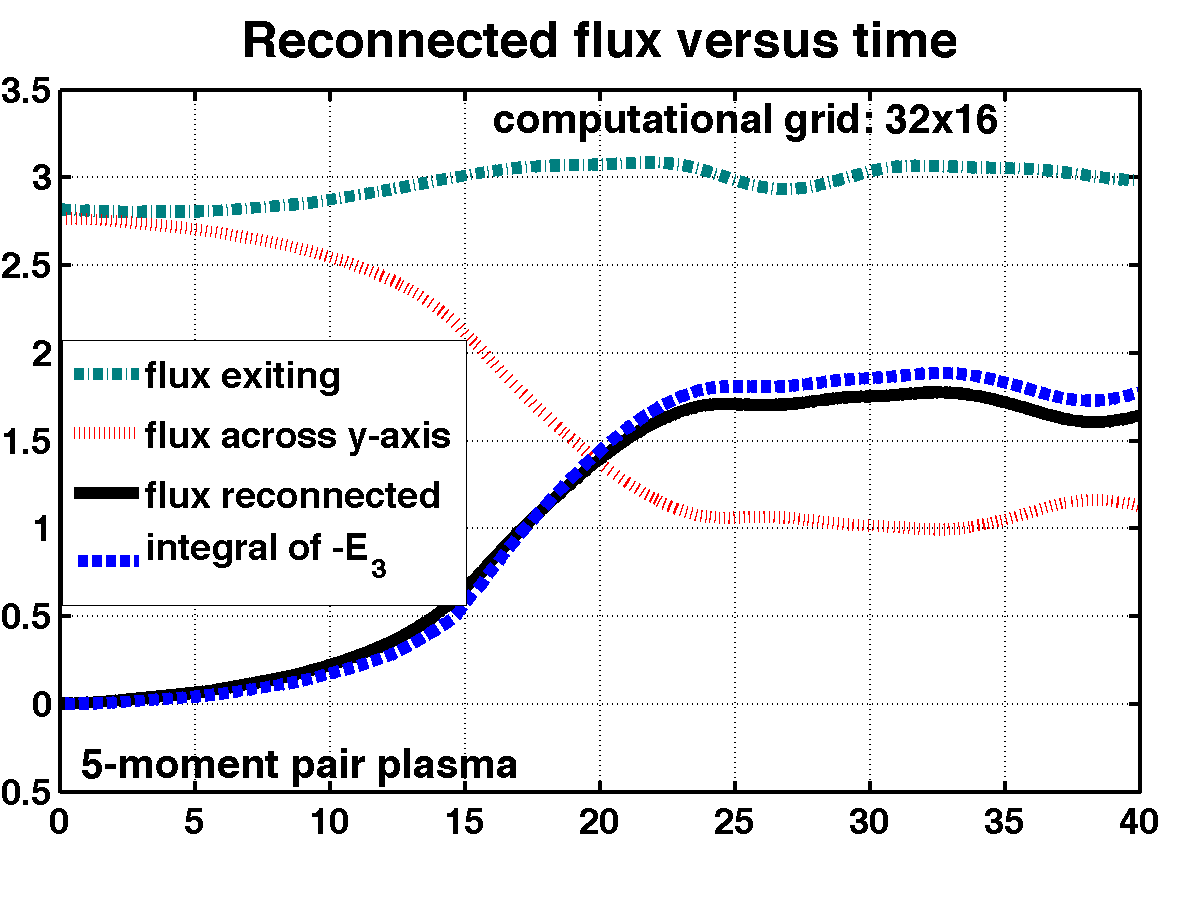}
  \includegraphics[width=3.25in, height=1.8in]{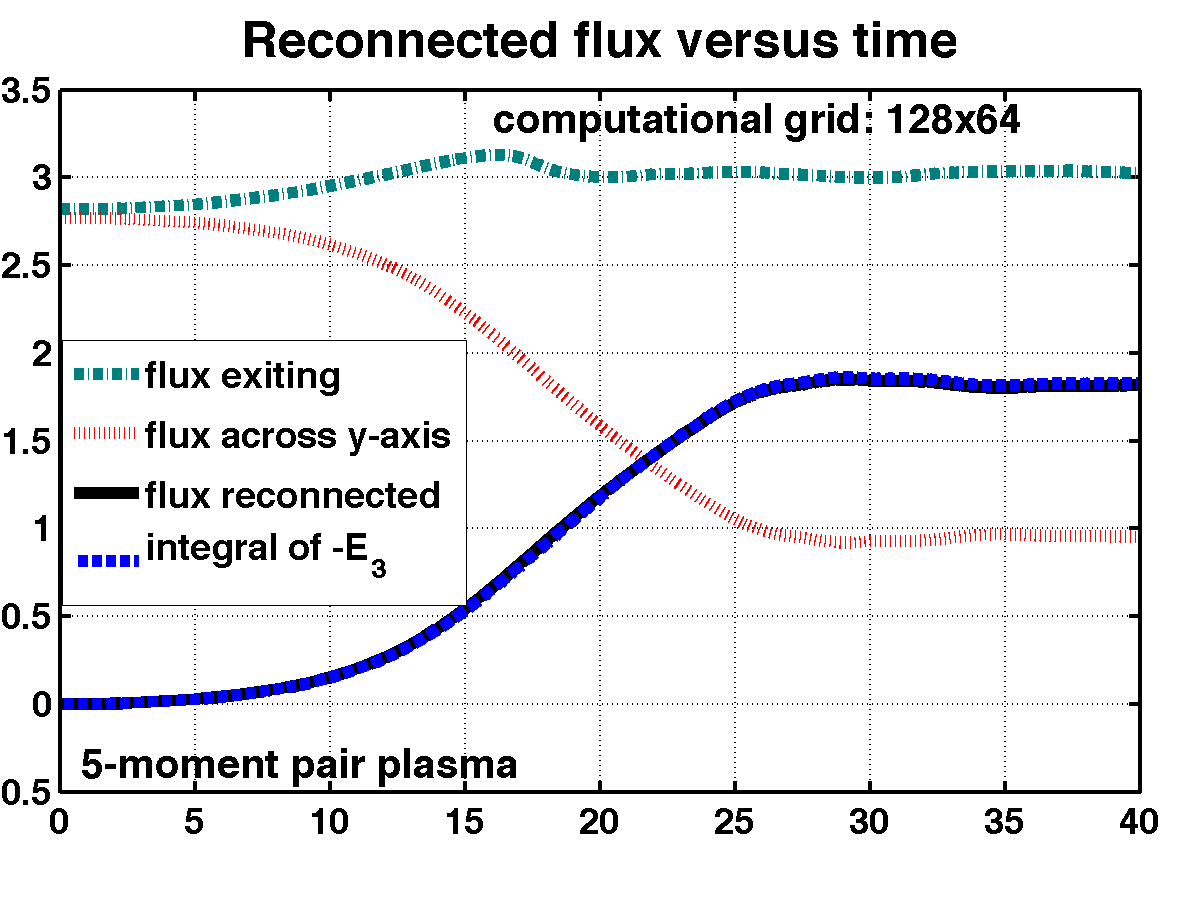}
  \caption{\small Examples of agreement of reconnected flux with minus the accumulation integral
    of the out-of-plane electric field at the origin for fine and coarse mesh.
    (For the fine mesh the two plots are exactly superimposed and so are indistinguishable.)
    Flux across the vertical axis represents the portion of field lines that have not reconnected.
    We obtained similar excellent agreement in all our simulations.}
  \label{f:recon_flux}
  \vskip2pt \hrule
\end{figure}

\begin{figure}[htbp!]
  \includegraphics[width=3.2in, height=1.6in]{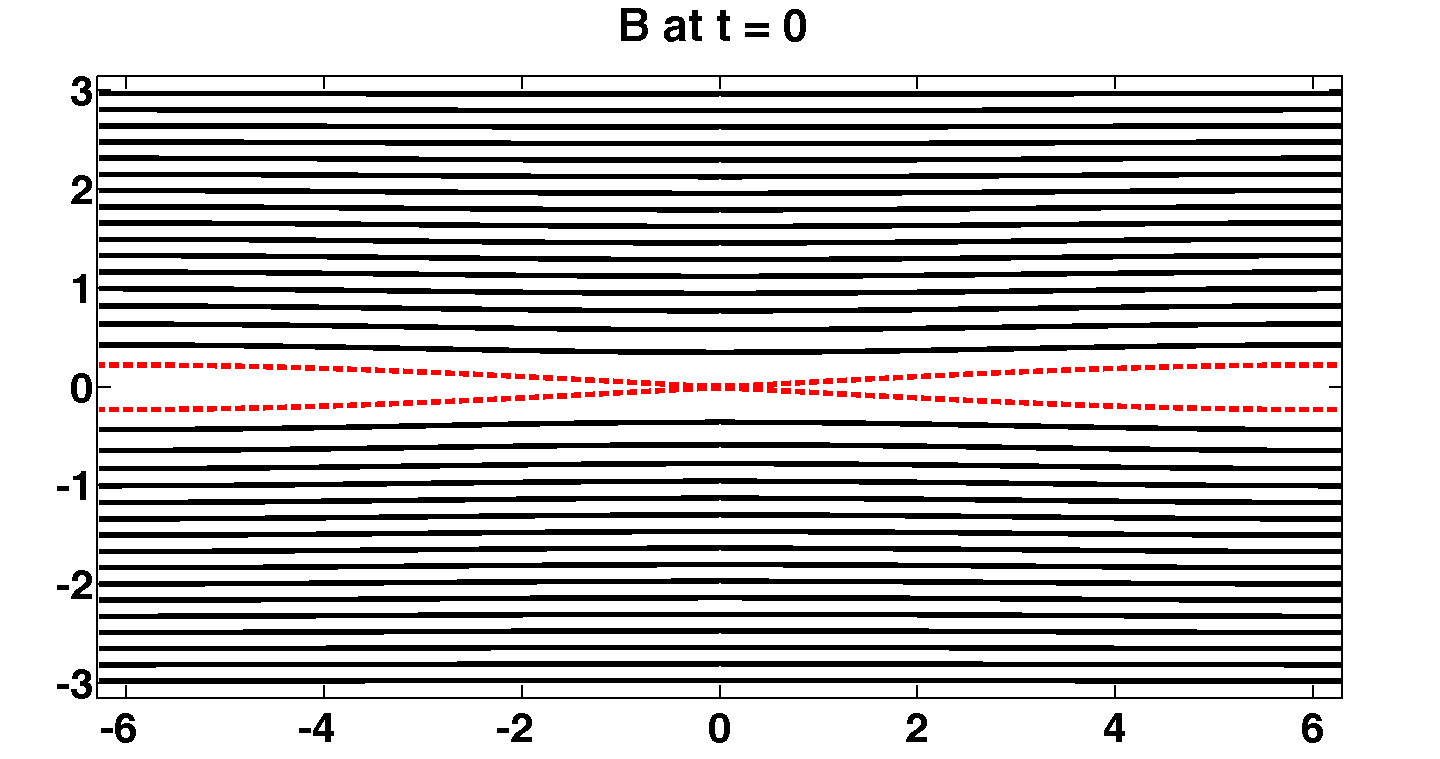}
  \includegraphics[width=3.2in, height=1.6in]{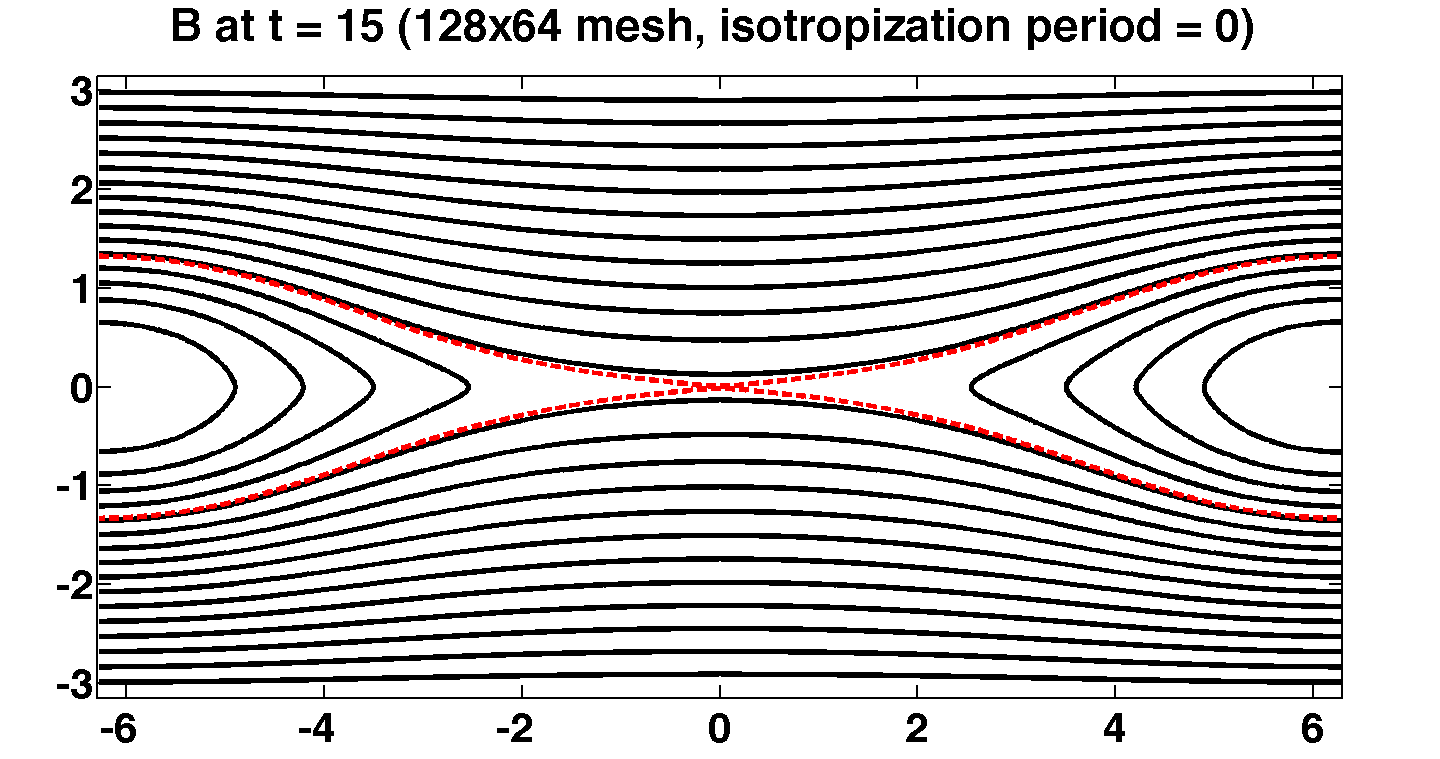}
  \includegraphics[width=3.2in, height=1.6in]{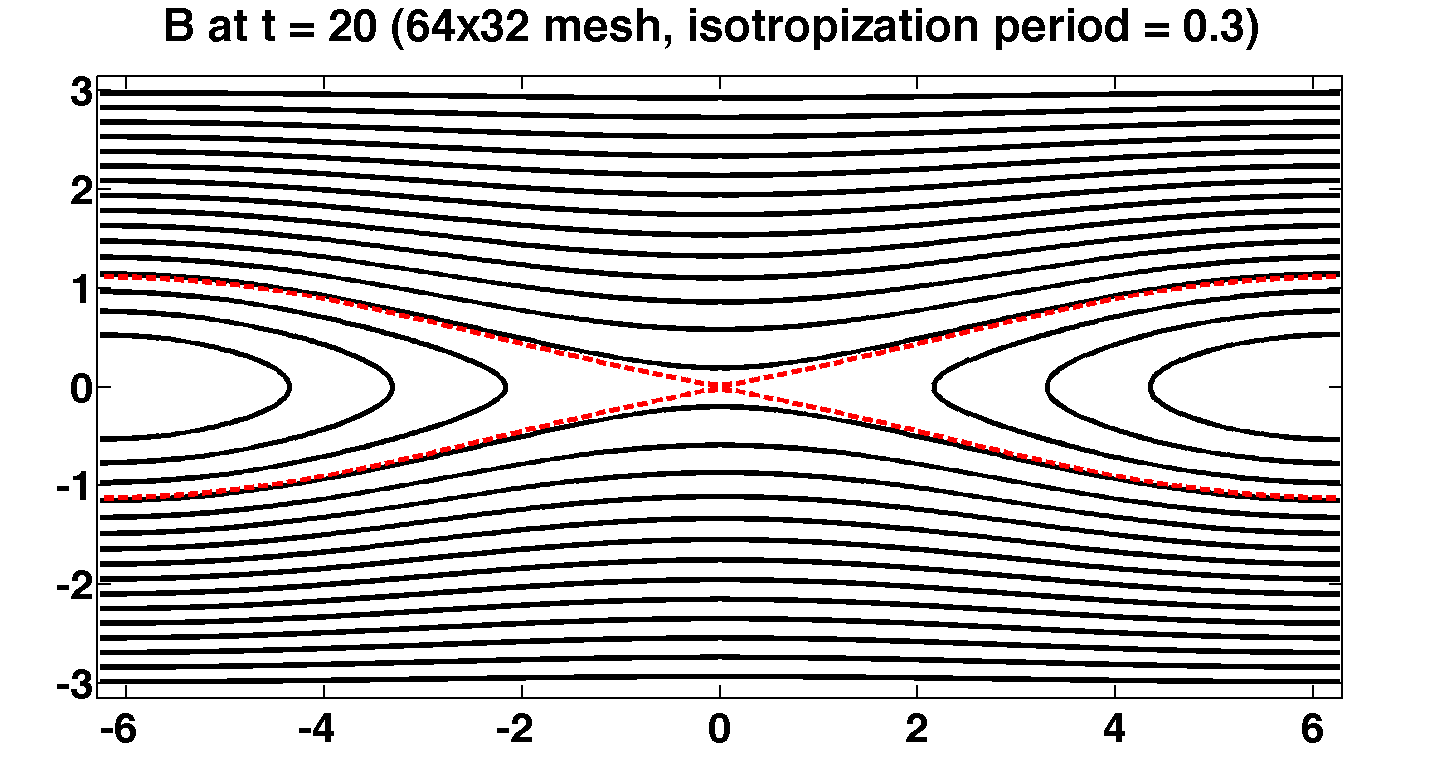}
  \includegraphics[width=3.2in, height=1.6in]{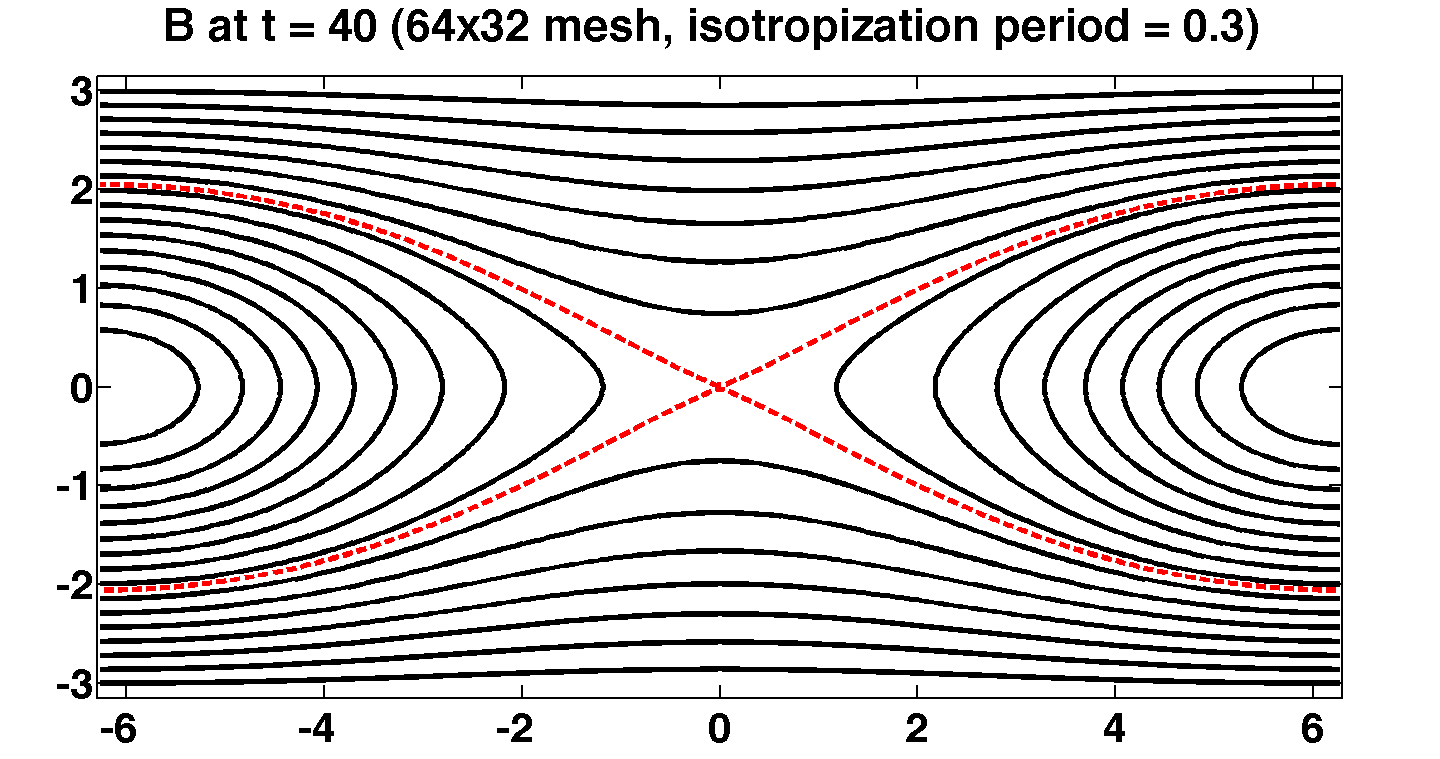}
  \includegraphics[width=3.2in, height=1.6in]{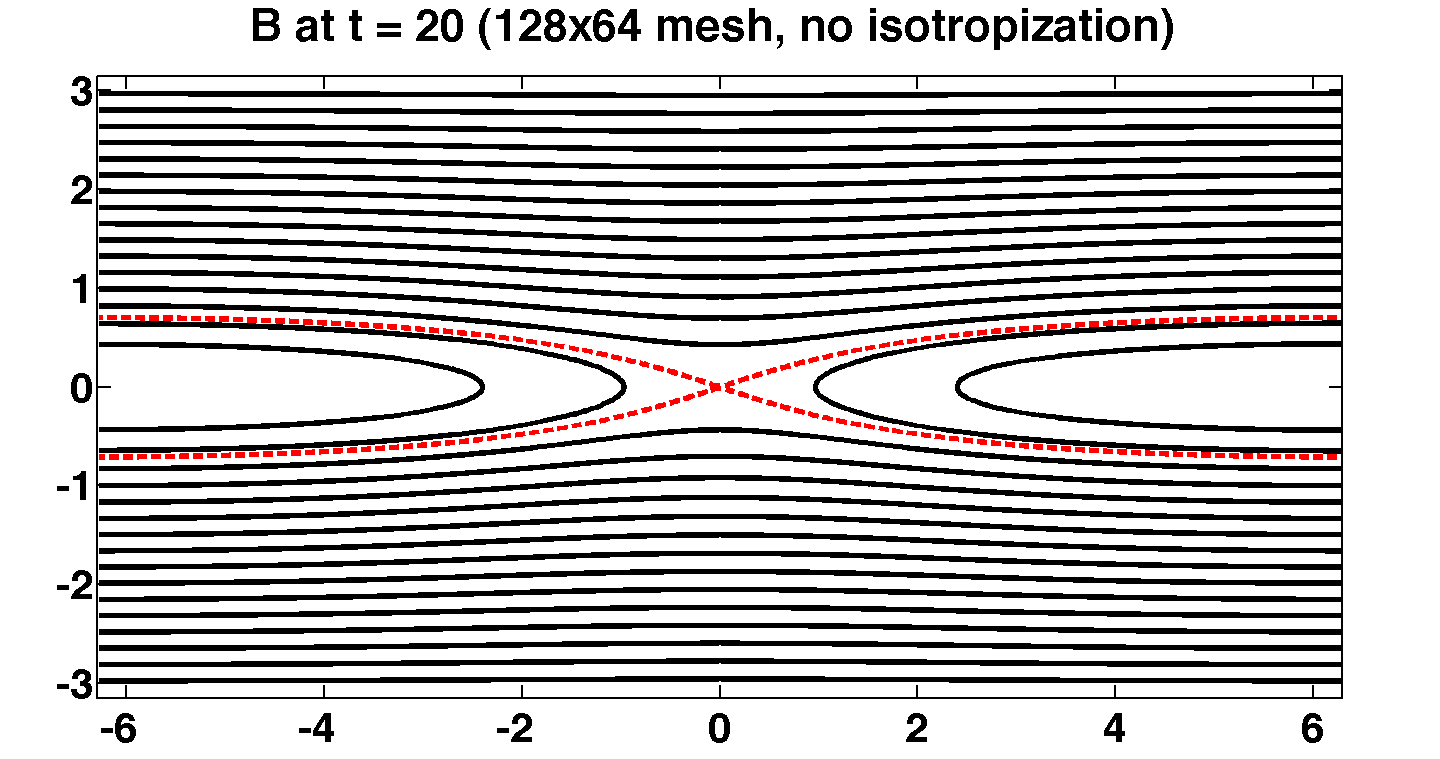}
  \includegraphics[width=3.2in, height=1.6in]{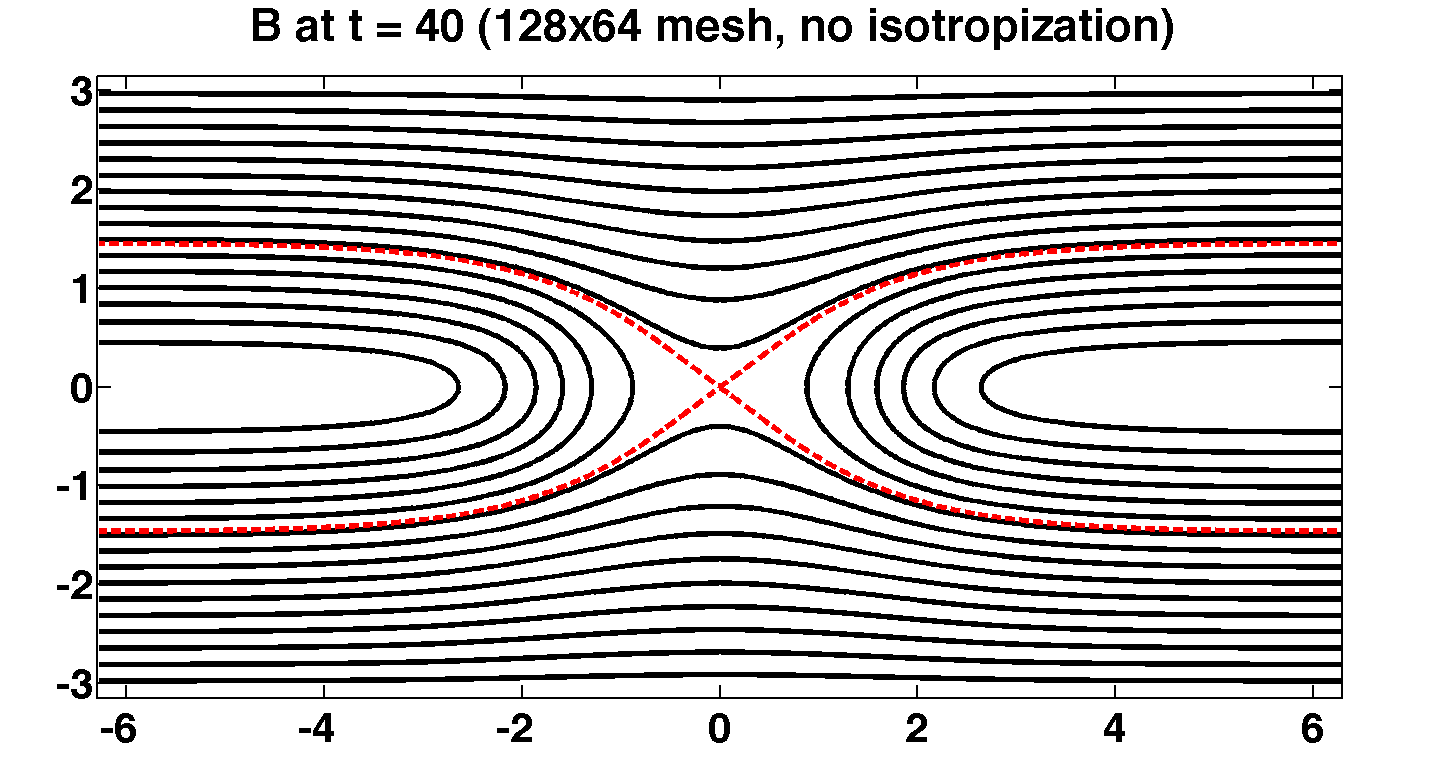}
  \caption{\small Snapshots of magnetic field lines.
    The reconnected flux is proportional to 
    (the change in) the number of field lines
    through the horizontal axis of symmetry, and the unreconnected
    flux is proportional to the number of field lines through the
    vertical axis of symmetry.}
  \label{f:magnetic_field_lines}
  \vskip2pt \hrule
\end{figure}

\vspace{-2mm}
{\bf Acknowledgements}
\vspace{-4mm}

I thank the Wisconsin Space Grant Consortium for their support.
In addition to my advisor, James Rossmanith,
I also thank Nick Murphy, Ellen Zweibel, and Ping Zhu
for helpful conversations.
\vspace{-5mm}

\bibliographystyle{plain}
\bibliography{/Users/evanjohnson/svn/bibliography/BigBib}

\end{document}